\documentclass[
]{elsarticle}

\usepackage[separate-uncertainty=true,
			exponent-product=\cdot,
			list-final-separator={, and },
			]{siunitx}
\DeclareSIUnit\Molar{M}
\DeclareSIUnit\rpm{rpm}
\usepackage[version=3]{mhchem}
\usepackage{amsmath}
\usepackage{amsfonts}
\usepackage{amssymb}
\usepackage{mathtools}
\usepackage[pdfauthor={Lennart H. Bosch, Pernille R. Jensen, Nico Striegler, Thomas Unden, Jochen Scharpf, Usman Qureshi, Philipp Neumann, Martin Gierse, John W. Blanchard, Stephan Knecht, Jochen Scheuer, Ilai Schwartz and Martin~B.~Plenio},
			pdftitle={Hierarchical Maximum Likelihood Estimation for Time-Resolved NMR Data}
			]
			{hyperref}

\hypersetup{
	colorlinks=true,
	urlcolor=black,
	linkcolor=black,
	citecolor=black,
}
\usepackage{graphicx}
\usepackage{booktabs}
\usepackage{caption}
\usepackage{titlesec}

\usepackage{enumitem}
\setlist{noitemsep}

\usepackage[
	top=2.5cm,
	bottom=2.5cm,
	left=15mm,
	right=15mm,
	footskip=1cm,
	headsep=0.75cm,
	columnsep=9mm,
]{geometry}
\usepackage[utf8]{inputenc}
\usepackage[T1]{fontenc}

\newcommand{\Cthirteen}{\ce{^{13}C}}
\newcommand{\Hone}{\ce{^{1}H}}
\newcommand{\CFum}{[1-\Cthirteen{}] fumarate}
\newcommand{\CPyr}{[1-\Cthirteen{}] pyruvate}
\newcommand{\CLac}{[1-\Cthirteen{}] lactate}

\newcommand{\iu}{{i\mkern1mu}}

\newcommand{\prob}{\ensuremath{P}}

\newcommand{\condprob}[2]{\prob{}\left( #1 \vert #2 \right)}

\newcommand{\fullstop}{\, .}
\newcommand{\ecomma}{\, ,}

\newcommand{\mvec}[1]{\vec{#1}}

\newcommand{\mmat}[1]{\mathbf{#1}}

\newcommand{\mop}[1]{\hat{#1}}

\newcommand{\expval}[1]{< #1 >}
\newcommand{\idmat}{\mathbb{I}}
\newcommand{\pardiff}[2]{\frac{\partial #1}{\partial #2}}

\def\true#1 {\hat{#1}}
\def\est#1 {#1_{\mathrm{est}}}
\def\transpose#1 {#1^{\intercal}}
\def\hc#1 {#1^{\dagger}}
\DeclareMathOperator{\rank}{Rank}

\newcommand{\sharetitle}{Hierarchical Maximum Likelihood Estimation for Time-Resolved NMR Data}
\title{\sharetitle{}}
\author[1]{Lennart H. Bosch\corref{cor1}}
\author[2]{Pernille R. Jensen}
\author[3]{Nico Striegler}
\author[3]{Thomas Unden}
\author[3]{Jochen Scharpf}
\author[3]{Usman Qureshi}
\author[3]{Philipp Neumann}
\author[3]{Martin Gierse}
\author[3,4]{John W. Blanchard}
\author[3]{Stephan Knecht}
\author[3]{Jochen Scheuer}
\author[3]{Ilai Schwartz}
\author[1,5]{Martin~B.~Plenio\corref{cor1}}

\cortext[cor1]{Corresponding author at: Institute of Theoretical Physics, Ulm University, Albert-Einstein-Allee 11, 89081 Ulm, Germany. \\ \href{mailto:lennart.bosch@uni-ulm.de}{lennart.bosch@uni-ulm.de}, \href{mailto:martin.plenio@uni-ulm.de}{martin.plenio@uni-ulm.de}}

\affiliation[1]{
	organization={Institute of Theoretical Physics, Ulm University},
	city={89081 Ulm},
	country={Germany},
	}
\affiliation[2]{
	organization={Center for Hyperpolarization in Magnetic Resonance, Technical University of Denmark},
	city={2800 Kgs. Lyngby},
	country={Denmark},
	}
\affiliation[3]{
	organization={NVision Imaging Technologies},
	city={89081 Ulm},
	country={Germany},
	}
\affiliation[4]{
	organization={Quantum Technology Center, University of Maryland, College Park},
	city={Maryland 20742},
	country={United States},
	}
\affiliation[5]{
	organization={Center of Integrated Quantum Science and Technology (IQST), Ulm University},
	city={89081 Ulm},
	country={Germany},
	}

\begin{document}
\begin{abstract}
Metabolic monitoring and reaction rate estimation using hyperpolarized NMR technology requires accurate quantitative analysis of multidimensional data scenarios.
Currently, this analysis is often performed in a two-stage procedure, which is prone to errors in uncertainty propagation and estimation.
We propose an approach derived from a Bayesian hierarchical model that intrinsically propagates uncertainties and operates on the full data to maximize the precision at minimal uncertainty.
In an analytic treatment, we reduce the estimation procedure to a least-squares optimization problem which can be understood as an extension of the Variable Projection (VarPro) approach for data scenarios with two predictors.
We investigate the method's efficacy in two experiments with hyperpolarized metabolites recorded with conventional high-field NMR devices and a micronscale NMR setup using Nitrogen-Vacancy centers in diamond for detection, respectively.
In both examples, the new approach improves estimates compared to Fourier methods and proves operational advantages over a two-stage procedure employing VarPro.
While the approach presented is motivated by NMR analysis, it is straightforwardly applicable to further estimation scenarios with similar data structure, such as time-resolved photospectroscopy.
\end{abstract}
\maketitle

\section{Introduction}
With applications in drug development and personalized medicine, metabolomics is of increasing importance in biomedical research and development~\cite{DeBerardinis2022a}.
Nuclear magnetic resonance (NMR) in combination with hyperpolarized samples has proven to be an effective means for the in vivo monitoring of metabolic conversions~\cite{Moco2022a,Ribay2023a,Peng2024a}.
Substrate hyperpolarization is of two-fold use in this context:
The signal strength is increased by factors up to \num{10000}~\cite{ArdenkjaerLarsen2003a} which reduces acquisition timescales below the second range and, therefore, granting temporal resolution for metabolic reactions~\cite{Day2007a,Gallagher2008a}.
Moreover, the signals from compounds of interest dominate the data and, therefore, are separable from background noise and other substances with ease.
Physiologically relevant metabolite concentrations are typically on the order of \qty{10}{\micro\Molar}~\cite{Bennett2009a} and, because of thermalization of the polarization, the signal continuously decreases in time, such that experiments operate close to detection limits.
Therefore, accurate reaction rate estimates and associated uncertainties are essential to high-quality research and development.
\par

Above referenced experiments are typically composed of a series of NMR sample snapshots over the course of an enzymatically induced reaction.
This leaves researchers with a two-dimensional dataset whose analysis is complicated and uncertainty propagation is susceptible to errors.
Over the years, multiple approaches for the quantitative analysis of this data have developed, but because of varying experimental context, there exists no generally applicable standard.
We provide a brief review of the most frequently used methods in both, time and frequency domain, along with relevant literature in the next section.
We proceed to present an approach based on a hierarchical Bayesian model, that is constructed from the expected relationships between signal intensities and, thereby, provides precise point estimates and accurate uncertainty estimation.
An analytic treatment of the model then reduces the parameter estimation procedure to a least-squares (LS) optimization problem, turning it into an easy to use tool.
The comparison of estimation results for the analysis of metabolic reaction data of HeLa cells recorded in a conventional NMR system and data from a J-coupling spectroscopy protocol executed on a micronscale NMR setup using Nitrogen-Vacancy (NV) centers for detection, demonstrates the new method's efficacy.
The new approach is motivated by, but not limited to parameter estimation in NMR experiments and can also be applied to various scenarios of LS estimation problems with systematically varying right-hand side~(RHS).

\begin{figure*}[!tbh]
\centering
\includegraphics[scale=1]{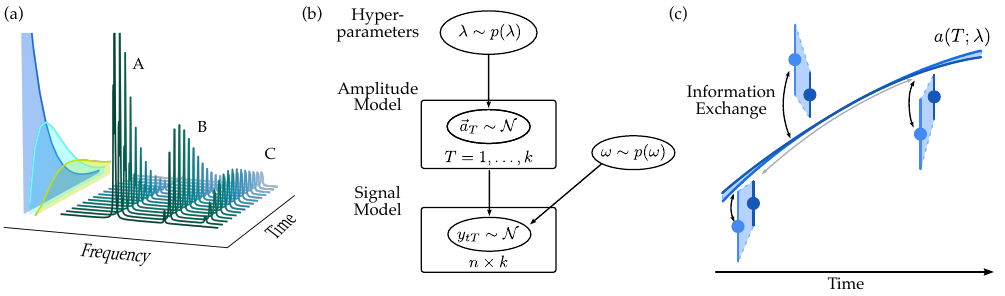}
\caption{(a) Schematic view of the two-dimensional structure of the data for a two-step conversion reaction. Shaded curves indicate the expected evolution of the three signal amplitudes. (b) Graphical representation of the hierarchical model. (c) Illustrative effects of the hierarchical approach onto individual amplitude estimates. Estimates are corrected using the second-level model while simultaneously reducing the uncertainty.}
\label{fig:data_sketch}
\end{figure*}

\subsection{Methods of Signal Quantitation in NMR} \label{sec:quantitation}
The most commonly used and widely accepted approach to intensity estimation is based on the area under the curve (AUC, also often denoted integral values) in the spectrum's real part, which can routinely be performed using any commercial NMR processing software such as TopSpin or MNova, as well as with community-maintained packages such as NMRPipe~\cite{Delaglio1995a}, nmrglue~\cite{Helmus2013a} or extensions to MATLAB.
To some extent, the outcome of this procedure depends on the operating user as phase correction and choosing the integration interval is often done by hand.
When signals overlap in frequency domain, the AUC estimation becomes particularly challenging and is typically done after baseline correction or by deconvolution.
Ultimately, however, the overlap implies correlation between intensity estimates (even if signal correction methods make it seem like they vanished), but they are often ignored in further processing.
Upon performing a consecutive analysis of intensities from different runs to, for example, estimate reaction rates, the associated uncertainties are not necessarily accurate because the intensity evolution of substrate and product is inherently correlated.
Moreover, the corrections applied to the data can distort the intensities, for example if signal intensities differ by multiple orders of magnitude, and parameter estimates are potentially inaccurate.
\par

A different branch of quantitation methods focuses on parameter estimation in time-domain, instead.
The compact representation of an NMR signal in frequency domain implies the time series to constitute a composition of a finite number of continuously damped sinusoids~\cite{Bretthorst1990a,Vanhamme1997a,Veen1988a}.
Assuming the frequency and damping parameters associated with the signal resonances known, estimates of the signal intensities are most commonly obtained from ordinary linear least-squares~(OLS) estimation, which by its relation to the principle of maximum likelihood~(ML) immediately provides intuitive uncertainty estimates ready for propagation into subsequent analyses~\cite[Chap.~2]{Weisberg2005a}, such as reaction rate estimation.
Methods for inference of frequency and damping parameters from the time-domain signal can be broadly divided into linear prediction-based methods~(LP)~\cite{Barkhuijsen1985a,Tang1988a,Gesmar1988a} and the direct fitting of nonlinear basis parameters~\cite{Macdonald1980a,Celmins1982a}.
Both methods share the assumption that observed data can be sufficiently well described as random numbers sampled from a probability distribution which is governed by a set of model parameters.
Linear prediction methods model the signal as stochastic autoregressive process, i.e., each signal sample occurs with a probability influenced by the previous samples~\cite{Marple2019a}.
Invoking the principle of ML and assuming additive white Gaussian noise (AWGN), the problem of finding the model parameters of a basis of damped sinusoids reduces to numerical root-finding of a polynomial, whose degree corresponds to twice the number of resonances assumed~\cite{Gesmar1988a}.
Implementations of LP methods are provided with most common software tools such as the ones named for the AUC analysis, and routinely also used for extrapolation of NMR signals and resolution enhancement~\cite{Zeng1989a}.
The direct fitting approach describes the maximization procedure of the likelihood distribution that describes the probability to observe the given data provided the underlying model parameters take a certain value.
Likewise assuming white Gaussian noise, the objective function for the optimization results in the commonly used (non-linear) least-squares~(NLLS) expression~\cite[Sec.~3.1]{Bishop2006a}.
The variable projection~(VarPro) procedure~\cite{Golub1973a,Kaufman1975a,Golub2003a,Pereyra2019a} (for applications in NMR, see \cite{Veen1988a,Boogaart1994a}) then eliminates the linear parameters from the optimization procedure, which can be retrieved upon subsequently invoking aforementioned OLS estimator.
The fitting procedure has been implemented in various software tools for NMR analysis, with its best known implementation AMARES~\cite{Vanhamme1997a} being deployed with the widely used software package jMRUI~\cite{Stefan2009a}.

Scenarios considered in this work exhibit systematic variation of signal composition over time, as schematically shown in fig.~\ref{fig:data_sketch}(a).
Both of the approaches to spectral analysis in time-domain can be generalized to multiple data vectors with varying linear coefficients (referred to as multiple RHS)~\cite{Marzetta1978a,Golub1979a,Kaufman1992a,Baerligea2023a}, for example, to fit multiple NMR signals with the same resonances at once or in the analysis of relaxation experiments~\cite{Gesmar1988b,Gesmar1989a,Chylla1995a}.
We typically expect the evolution to follow a smooth second level model, indicated by the curves on the left side of fig.~\ref{fig:data_sketch}(a) and, thus, regard the RHS as structured.
If specified accordingly, a hierarchical Bayesian model can exploit this structure in the parameter estimation procedure while intrinsically respecting correlations and naturally propagating uncertainties directly into estimates of second level parameters.
The specific model tailored to the scenarios considered in this work is shown in the graph in fig.~\ref{fig:data_sketch}(b).

The use of Bayesian models require an adjusted interpretation of uncertainties and naturally allow to incorporate additional prior knowledge into the estimation procedure.
For an introduction to Bayesian parameter estimation, we refer to a series of publications by Bretthorst~\cite{Bretthorst1990a,Bretthorst1990b,Bretthorst1990c}, who pioneered the Bayesian treatment of NMR signals\footnote{A more general introduction into Bayesian parameter estimation is provided with the book by Kruschke~\cite{Kruschke2015a}.}.
There also exist generalizations of Bretthorst's work to multidimensional data scenarios in NMR~\cite{Chylla1993a}, albeit not making use of the  hierarchical estimation structure.
While today estimation from the posterior distribution is commonly performed with Markov-Chain Monte Carlo (MCMC) sampling~\cite[Ch.~7]{Kruschke2015a}, we resort to analytical marginalization to simplify the problem and, afterwards, efficiently derive point estimates from the well-behaved posterior distribution via numerical optimization, instead.

If LP–based methods and NLLS fitting are based on the same underlying signal model of exponentially damped sinusoids, their estimates of frequency and damping parameters are known to agree asymptotically under large sample size and under identical assumptions on noise statistics, model order, and signal-to-noise ratio~(SNR)~\cite{Osborne1995a,Wu1981a}.
As a consequence, the resulting intensity estimates obtained from both approaches are consistent, rendering it sufficient to choose either of them, in addition to the AUC results, as a reference for comparison with the newly introduced hierarchical approach.
Upcoming discussions will reveal a close relation between the newly introduced approach and NLLS in eq.~\eqref{eq:neglogl}, making it the preferred choice; however, a change in the specific method used for spectral estimation does not affect the results presented in this work.

\section{Theory}
\label{sec:theory}
Probabilistic approaches to parameter estimation share the fundamental assumption that observed data can be sufficiently well described as random numbers sampled from a probability distribution that depends on true value of generic parameters of interest $\true{\theta} $, marked by the hat.
Thus, the observed value $y_t$ at time point $t$ is treated as a sample from the so called likelihood distribution $\prob \Big( y \vert \true{\theta} , t\Big) $.
To estimate the unknown parameter $\theta $ from the principle of ML, one constructs an explicit expression of the probability distribution, plugs in the observed sample $y_t$ and maximizes with respect to $\theta$.
The value obtained for $\theta$ then corresponds to the model parameter under which it is most likely to observe the data $y$.
The distribution's width with respect to $\theta $ is then also typically used as a measure of uncertainty.

For most problems, a single sample is insufficient to determine the unknown model parameters, such that the probability distribution must be generalized to a data vector as $\prob \big( \mvec{y} \vert \theta \big) $, with the explicit dependence on the time points being dropped here to simplify the notation.
The distribution's explicit expression can take vastly different shapes and is based on the type of noise expected, and whether it is correlated in time.
Under the simplest and most widely assumed noise model AWGN, the samples are assumed to be independently sampled from a Gaussian distribution whose mean value is determined by the model value
$y_t \ \sim \mathcal{N}(f(\theta ;t),\sigma^2)$
or equivalently
$ y_t = f( \theta ;t) + \varepsilon_t$
with $\varepsilon_t \sim \mathcal{N}(0,\sigma^2)$ independently, identically distributed (i.i.d.) following a Gaussian (normal) distribution.
In this simple case, the generalization to $n$ samples is given by the product of the individual probabilities as
\begin{align}
\prob ( \mvec{y} \vert \theta) = \frac{1}{\sqrt{2\pi\sigma^2}^n}\exp( \frac{1}{2\sigma^2}\Vert \mvec{y} - \mvec{f}(\theta) \Vert^2) \ecomma
\end{align}
where components of $\mvec{f}$ correspond to the model evaluated at all available $t \in \lbrace t_1, t_2, \dots , t_n \rbrace$.

In contrast to the ML approach, the Bayesian approach to parameter estimation aims to derive statements from the posterior distribution $\prob \big( \theta \vert \mvec{y} \big) $, instead.
This allows for the more natural interpretation of what is the most probable parameter value for $\theta$ given the data $\mvec{y}$.
The posterior distribution is related to the likelihood distribution via Bayes' theorem~\cite{Bayes1763a}
\begin{align}
\prob \big( \theta \vert \mvec{y} \big) = \frac{\prob \big( \mvec{y} \vert \theta \big) \cdot \prob (\theta)}{\prob (\mvec{y})} \, , \label{eq:bayestheorem}
\end{align}
with $\prob (\theta)$ being referred to as the prior distribution and $\prob (\mvec{y})$ merely representing a normalization constant here.
The prior distribution summarizes the information about the parameter that was gained independently from the current experiment.
It can, for example, provide hard limits on parameter values or even incorporate additional knowledge of the parameter if available~\cite[Chap. 5]{Kruschke2015a}.
If one wants to mimic the case of little to no prior knowledge, it can be chosen sufficiently flat over the region of interest not to distort the estimation given $\mvec{y}$.
Beyond, the prior distribution can be used to connect the experiment yielding $\mvec{y}$ to previous or similar experiments to perform a joint analysis\footnote{Details of the explicit formulation may vary vastly and are still under scientific debate~\cite{Mikkola2024a}.}.
Assuming $k$ runs in total, the parameter $\theta$ of each individual run is then, for example, treated as intermediate parameter that can be used to determine a set of so-called hyperparameters $\lambda$.
Formally, the posterior distribution in eq.~\eqref{eq:bayestheorem} is extended upon substitution of the prior distribution $\prob (\theta)$ by $\prob (\theta \vert \lambda) \cdot \prob (\lambda)$ to
\begin{align}
\prob \big( \theta, \lambda \vert \mvec{y} \big) \varpropto \lbrace \prob \big( \mvec{y} \vert \theta \big) \cdot \prob (\theta \vert \lambda) \rbrace \cdot \prob ( \lambda ) \fullstop
\end{align}
Assuming AWGN, the generalization to $k$ runs of the same experiment, each with its own value of $\theta$ is composed of a product of individual probabilities as
\begin{align}
\prob \big( \lbrace \theta \rbrace, \lambda \vert \lbrace \mvec{y} \rbrace \big) \varpropto \prod\limits_{l=1}^{k} \lbrace \prob \big( \mvec{y}_l \vert \theta_l \big) \cdot \prob (\theta_l \vert \lambda) \rbrace \cdot \prob ( \lambda ) \fullstop \label{eq:generic_hm}
\end{align}
The resulting model is then typically referred to as hierarchical model and can be visualized as in figure~\ref{fig:data_sketch}(b).
The posterior now provides means for a joint estimation of parameters $\theta_l$ for each run, as well as for joint estimation of hyperparemeters $\lambda$ connecting multiple runs of the same experiment.

In the scenario addressed in this work, the NMR system dynamics are attributed to two dimensions whose time parameters are referred to with $t$ for the direct and $T$ for the indirect dimension.
For fixed $T$, dynamics on the faster timescale $t$ are described by a model composed by a linear superposition of generally nonlinear basis functions as
\begin{align}
\mvec{f}_T (\mvec{a}_T, \omega) = \Phi(\omega)\cdot \mvec{a}_T \ecomma \label{eq:abstract_model}
\end{align}
as commonly used in the description of NLLS problems~\cite{Golub1973a,Kaufman1975a,Kaufman1992a,Baerligea2023a}.
The basis functions are evaluated for all discrete $t$ at which data is recorded and the resulting vectors are organized into columns of the matrix $\Phi$.
For an NMR experiment, the $j$-th basis function typically corresponds to the damped sinusoid signal of the $j$-th resonance that is observed with amplitude given by the $j$-th element of the vector $a_T$.
Consequently, components of the model $\mvec{f}_T$ then represent the sum of all resonances evaluated for all discrete time points.

Model parameters that were previously summarized as $\theta$ are now explicitly separated into frequencies $\omega$ and amplitudes $\mvec{a}_T$.
Assuming the resonance frequencies to remain invariant between different runs, we can assume the same parameters $\omega$ for different runs of a the same experiment.
The amplitude parameters $\mvec{a}_T$, however, are expected to vary on the larger timescale $T$ and follow a second level model  $\mvec{a}(T;\lambda)$.
Thus, data recorded at discrete $T$ is expected to vary only in the composition of individual basis functions or, equivalently, resonances and the matrix $\Phi$ is being reused.
Following the Bayesian approach introduced above, we replace parameter $\theta_l$ in eq.~\eqref{eq:generic_hm} by $\omega,\mvec{a}_T$ (also changing the summation index from $k$ to $T$) and take into account, that frequencies $\omega$ are also shared by all runs.
Under assumption of samples being independently distributed such that the collective probability distribution can be written as product of the probability for each individual sample, a formal description of the expected posterior distribution for the full two-dimensional data reads
\begin{align}
\condprob{\mmat{A},\omega,\lambda}{\mmat{Y}}
\varpropto \prod\limits_T \lbrace \condprob{\mvec{y}_T}{\mvec{a}_T,\omega} \cdot \condprob{\mvec{a}_T}{\lambda}\rbrace \cdot \prob{} (\omega) \cdot \prob{}(\lambda) \ecomma
\label{eq:hierarchical_posterior}
\end{align}
where matrices $\mmat{A}$ and $\mmat{Y}$ are composed by columns of $\mvec{a}_T$ and $\mvec{y}_T$ at different $T$, respectively.
This distribution now serves as basis for estimation of parameter values for all free parameters, including $\lambda$ which can, for example, represent free parameters of a kinetic model that describes changes in signal intensities of different resonances over time.

Typical experiments to resolve enzymatic reactions or relaxation experiments often consist of more than $100$ runs, each with multiple resonance amplitudes.
Thus, the posterior distribution in eq.~\eqref{eq:hierarchical_posterior} easily features multiple hundred free parameters that are potentially correlated, making the parameter estimation procedure very costly, independent of the exact algorithm and posterior distribution used.
If we are mainly interested in the hyperparameters $\lambda $ that, for example, correspond to reaction parameters, we can try and analytically marginalize the distribution over all elements of $\mmat{A}$ by integration, yielding a distribution whose number of free parameters remains independent of the data size:
\begin{align}
\condprob{\omega,\lambda}{\mmat{Y}} = \int\limits_{-\infty}^{\infty} \condprob{\mmat{A},\omega,\lambda}{\mmat{Y}} \ \mathrm{d}\mmat{A} \fullstop \label{eq:generic_marginalization}
\end{align}
We proceed and define the marginal likelihood as
\begin{align}
\mathcal{L}(\omega,\lambda) \varpropto \frac{\condprob{\omega,\lambda}{\mmat{Y}} }{\prob{} (\omega) \cdot \prob{}(\lambda)} \ecomma \label{eq:general_hierarchical_likelihood}
\end{align}
to decouple upcoming discussions from the choice of prior distributions.
The likelihood can then be treated using the principles of maximum-likelihood estimation to obtain estimates of $\omega$ and $\lambda$ as well as associated uncertainties.
Due to its origin from a hierarchical model, we will refer to this method as hierarchical maximum-likelihood~(HML) approach to parameter estimation.
Upon multiplication with priors for $\omega$ and $\lambda$, we retrieve parameter estimation by Bayesian principles, again.

Under the assumption of AWGN for data samples in $\mmat{Y}$, the full data matrix describing multiple FID runs is modeled as
\begin{align}
\mmat{Y} = \Phi(\omega) \cdot \mmat{A} + \mmat{E} \ \ \text{with} \ \ E_{ij} \sim \mathcal{N}(0,\sigma^2) \ \text{i.i.d.} \fullstop
\end{align}
In addition, choosing a Gaussian conditional probability $\condprob{\mvec{a}_T}{\lambda}$ with covariance $\sigma^2 (\transpose{\mmat{\Phi}} \mmat{\Phi})^{-1} $ motivated by the above noise model and known as Zellner's g-prior with $g=1$~\cite{Zellner1986a}, the analytical marginalization in eq.~\eqref{eq:generic_marginalization} and plugging the result into eq.~\eqref{eq:general_hierarchical_likelihood} yields the compact expression
\begin{align}
- \ln \mathcal{L}(\omega,\lambda)
\varpropto \frac{1}{2\sigma^2} \cdot \lbrace \frac{1}{2} \Vert \mmat{Y} -\Phi \Phi^\dagger \mmat{Y} \Vert_2^2 + \frac{1}{2} \Vert \mmat{Y}-\Phi \mmat{A}(\lambda) \Vert_2^2 \rbrace \ecomma
\label{eq:neglogl}
\end{align}
where the subscript 2 indicates the use of the Frobenius norm (squared sum of all matrix elements) and $\Phi^\dagger=(\transpose{\Phi} \Phi)^{-1} \transpose{\Phi} $ denotes the Moore-Penrose-Pseudoinverse~\cite{BenIsrael2003a}.

The main result of the theoretical treatment in this work eq.~\eqref{eq:neglogl} represents an equally weighted average of two terms:
The first term is identical to the VarPro objective function~\cite{Golub1973a,OLeary2013a} trivially generalized to multiple dimensions~\cite{Golub1979a} and the second term computes the square distance between data sample and the model when imposing a perfect propagation of the higher level model to the data samples.
Thus, it can be viewed as an extension of the VarPro approach for NLLS problems with a structured RHS.
The additional prefactor of $1/2\sigma^2$ is linked to the distribution's width and, thus, ensures accurate uncertainty estimates, as discussed in \ref{app:marginalization}.
For the full expressions of probability distributions used and the derivation of the result~\eqref{eq:neglogl} refer to \ref{app:hm_motivation} to \ref{app:HML}.

In this treatment, the problem generally does not separate into independent estimation problems for lower level parameters $\omega$ and upper level counterparts $\lambda$, indicating correlation between estimates for both parameter species.
If derived from the objective function's curvature, represented by the Hessian matrix, the estimation uncertainty depends linearly on the noise standard deviation $\sigma$ and, thus, scales reciprocally to the SNR.
Moreover, if $\mmat{A}(\lambda)$ perfectly coincides with the ordinary LS estimate of $\Phi^\dagger \mmat{Y}$ of $\mmat{A}_T$, values and curvature w.r.t. $\omega$ are the same as from the standard NLLS expression.
Reorganization of the individual sum terms into a single vector reduces the problem to a LS optimization problem, such that we can write down the Jacobian Matrix or find the optimum using a number of well known optimization routines.
For a detailed discussion and an expression of the Jacobian refer to \ref{app:LS_form}.

If explicitly interested in estimates of individual amplitudes $\mmat{A}$, the Bayesian recipe instructs us to plug estimates of $\omega$ and $\lambda$ back into the full distribution $\condprob{\mmat{A},\omega,\lambda}{\mmat{Y}}$ and maximize with respect to $\mmat{A}$,
leaving us for flat priors with
\begin{align}
\est\mmat{A} = \frac{1}{2} \left[ \Phi^\dagger(\est\omega{}  ) \mmat{Y} + \mmat{A}(\est\lambda{}  ) \right] \label{eq:hierarchical_ampl_estimate}
\end{align}
and covariance matrix
\begin{align}
\Sigma_{\mvec{a}} = \frac{\sigma^2}{2} ( \Phi^\intercal{}(\est\omega{}  ) \cdot \Phi(\est\omega{}  ) )^{-1} \label{eq:hierarchical_ampl_covariance}
\end{align}
identical for all $T$.
Thus, the hierarchical approach corrects the ordinary NLLS estimates to the average of NLLS estimates and the model in the indirect dimension while reducing the covariance by a factor of $1/2$.
This effect is visualized schematically in fig.~\ref{fig:data_sketch}~(c) and its derivation more thoroughly discussed in \ref{app:ampl_est}.
\par

The inclusion of prior knowledge on the signal in the direct dimension is possible in two ways:
Either specify the prior distribution $\prob(\omega)$ accordingly, which adds a regularization term in logarithmic space.
Alternatively, we can include constraints into the optimization procedure, which for example fix differences between resonances.
Taking the implementation of AMARES~\cite{Vanhamme1997a} as role model, one can replace the corresponding parameters in the functional or resort to standard regularization terms typically used for constrained NLLS problems.
To include prior knowledge for estimation of parameters $\lambda$, the same rules apply.

\section{Results}
\subsection{Reaction Rate Estimation of HeLa Cells} \label{sec:reaction_rate_est}
Since the advent of hyperpolarization as a tool for NMR, e.g., by use of dissolution Dynamic Nuclear Polarization~\cite{ArdenkjaerLarsen2003a} or more recently Para-Hydrogen Induced Polarization~\cite{Bowers1986a,Bowers1987a,Reineri2015a,Dagys2024a}, the study of metabolic pathways and reaction rates has evolved into a standard procedure in biochemical research~\cite{Ribay2023a}.
In the typical setup, a series of FID experiments with small flipping angles is recorded over timescales of the targeted reaction.
The reaction dynamics observed in the indirect dimension (associated with $T$) are slow compared to dynamics in the direct dimension (associated with $t$) and dominantly manifest in the variation of signal amplitudes.
Moreover, the observed noise can be well approximated by AWGN, such that these experiments turn out as compelling use-case of the HML approach for parameter estimation.
We apply the analysis scheme introduced in Section~\ref{sec:theory} to a series of FID experiments with hyperpolarized \CPyr{} recorded over the course of \qty{240}{\s} (see figs.~S1-S3 of the supplementary material) and compare the reaction rate estimates to results from the AUC approach as described in Section~\ref{sec:quantitation}.
To further investigate the specific behavior of these approaches and point out pitfalls and flaws in the two-step procedure using VarPro, we perform a simulation based on parameters retrieved from aforementioned experiment, granting precise control over the noise level, and perform an identical analysis.

\begin{figure}[tb]
\centering
\includegraphics[scale=1]{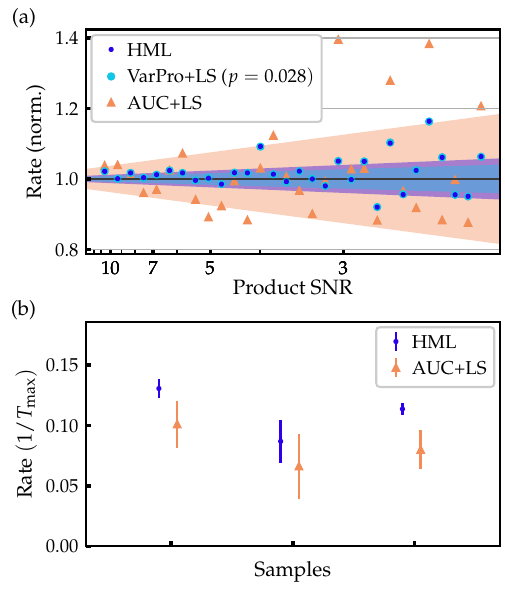}
\caption{(a) Reaction rate estimates from three different approaches derived from simulated data for different realizations of noise. HML stands for hierarchical maximum likelihood, VarPro+LS estimates come from a two-step procedure in time-domain using VarPro followed by LS fitting of the kinetic model and AUC denotes the area under the curve of the spectrum's real part also followed by a fit of the kinetic model. Shaded areas indicate uncertainty range. Agreement of uncertainty estimates, indicated by the shady areas, with the CRB is associated with the shaded area hosting about two thirds of the data points. Disagreement is observed only for VarPro+LS with $p=\num{0.028} < \num{0.05}$. (b) Reaction rate estimates derived from experimental data obtained with three different samples of HeLa cells with hyperpolarized \CPyr{}. Uncertainty estimates are reduced \num{2}- to \num{5}-fold. The systematic discrepancy can be explained by flaws in the AUC approach when signal intensities are observed on different orders of magnitude.}
\label{fig:results_plot}
\end{figure}

Conventional NMR devices record the magnetization in two orthogonal dimensions and the data is combined into a complex-valued signal representing the magnetization quadrature.
A well-established description of the data in time-domain is given by basis functions of complex exponentials with Lorentzian lineshape~\cite{Bretthorst1990a,Vanhamme1997a,Veen1988a}
\begin{align}
\phi_j(t;\omega,k,\varphi) = \exp(\iu \omega_j t - \eta_j t + \varphi_j) \label{eq:basisfunc}
\end{align}
with frequency $\omega_j$, decay rate $\eta_j > 0$ and phase $\varphi_j \in [0,2\pi)$ associated with the $j$-th resonance are assumed identical for each FID run.
Because the information available from the kinetic model is limited to real-valued amplitudes $\mvec{a}$, the real and imaginary parts of the basis functions are concatenated to form the real-valued matrix $\Phi(\omega,\eta,\varphi)$ used to compute the signal model as in eq.~\eqref{eq:abstract_model}.
Conceptually, the complex phase factor in the basis function~\eqref{eq:basisfunc} could be accounted for via complex-valued amplitudes, instead, allowing for variation in the phase between individual FID runs.
This, however, would also require an extension of the amplitude model to the complex domain and while mathematically possible, this procedure is not sufficiently backed by the information available, namely the real-valued model for signal intensities.
For a more extensive discussion on the use of complex-valued basis functions and amplitudes as well as other choices for the prior refer to \ref{app:polar_coords}.

Changes in signal intensity due to enzymatically induced conversion and thermalization of the polarization observed in the indirect dimension are described by a simple first-order conversion model~\cite{Knecht2021a}:
\begin{align}
\dot{\mvec{a}}(T;\lambda) =
\begin{pmatrix}
-\kappa_\mathrm{P} - k & 0 \\
k & -\kappa_\mathrm{L}
\end{pmatrix}
\cdot
\mvec{a}(T;\lambda)
\fullstop
\end{align}
Reaction rate $k$, thermalization rates $\kappa_\mathrm{P}$ and $\kappa_\mathrm{L}$ and initial values $\mvec{a} (0;\lambda)=\transpose{(P_0 \ L_0)} $ for pyruvate and lactate, respectively, are left variable for the estimation.
\par

Using data values provided in \ref{app:hela_exp}, a numerical experiment with simulated signals is performed to confirm the HML method's expected behavior and compare the results with two other analysis approaches:
\begin{itemize}
\item Intensity estimation from AUC (after manual phase correction) and a consecutive LS fit of the amplitude model;
\item VarPro approach on timescale $t$ and consecutive LS fit of the amplitude model.
\end{itemize}
The simulation is performed with different realizations of the noise, once for each of \num{30} equidistantly distributed values of $\sigma$.
Estimation results of the reaction rate are plotted in fig.~\ref{fig:results_plot}(a) over the reaction product's maximal SNR observed in the FID run with the maximal intensity.
To declutter the image, the estimated uncertainty each is indicated by the shaded area drawn around the mean estimated value.
While mean estimates of all three methods agree with the true value, the AUC+LS approach's uncertainty is approximately \SI{50}{\percent} larger as by HML, which again is about \SI{40}{\percent} larger than for VarPro+LS.
The systematic difference in uncertainty between VarPro+LS while point estimates appear almost identical hints at an error in the uncertainty estimation procedure.
ML estimators are known to be efficient estimators in the sense that they saturate the Cramér-Rao lower bound~(CRB) for estimator variance and meaningful uncertainty estimates are expected to be consistent with estimator variance.
Thus, upon agreement of the estimated uncertainty with the CRB, we expect \SI{68.3}{\percent} of data points to asymptotically lie in the corresponding shaded area.
Treating data data points in fig.~\ref{fig:results_plot} as samples from a binomial distribution and assuming agreement with the CRB, the hypothesis test yields $p=\num{0.028} < \num{0.05}$ for VarPro+LS, indicating statistically significant disagreement between the CRB and the computed uncertainty estimate.
For the other methods, no disagreement was found.
The underestimation is a consequence of neglecting correlations in the uncertainty propagation, which can be corrected for by inserting the full covariance matrix in the LS objective function resulting in uncertainty estimates that approximately coincide with HML uncertainty estimates.
Because of the incorrectly estimated uncertainty, the VarPro+LS approach will not be considered further in its current implementation.
Investigation of the scaling of the estimated uncertainty with respect to further parameters, such as lineshapes and the kinetic rate, reveals similar relative scaling behaviour for all investigated methods.

The reaction rate estimates derived from experimental data of three different samples with HeLa cells are shown in fig.~\ref{fig:results_plot}(b).
While qualitatively exhibiting the same variation, the systematic difference between the two estimates is expected to stem from the manual correction of the spectral data before the AUC analysis and the finite integration intervals.
Moreover, we observed a systematic discrepancy between the data and model which leads to a slight overestimation of the uncertainty when computed from the residuals (see figs.~S7-S9 of the supplementary material).
Showing the same discrepancy to the model, the HML leads to an uncertainty reduction by a factor of about \numrange{2}{5} in the considered scenario.
The overall estimates are all in the same order of magnitude, thereby implying consistency in the estimation procedure.
For details on the data processing and estimation procedure, refer to \ref{app:hela_exp}.
Plots of the signal residuals are provided in figs.~S7-S9 of the supplementary material.
\par

\subsection{J-coupling Spectroscopy in Micronscale NMR using NV Centers} \label{sec:microNMR}

\begin{figure*}[bt]
\centering
\includegraphics[scale=1]{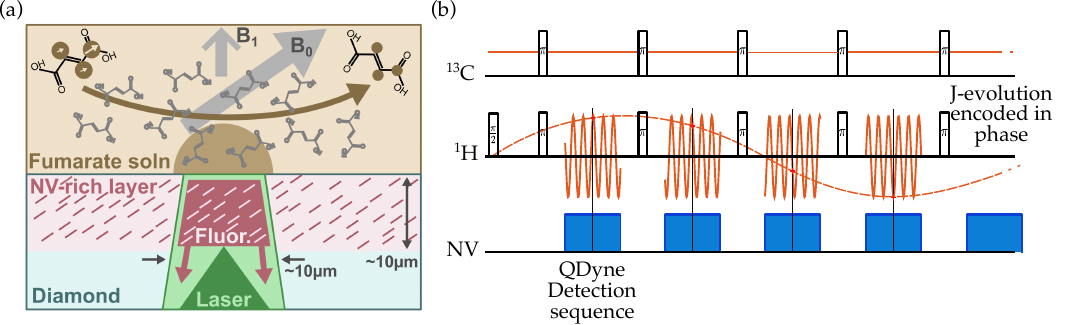}
\caption{
(a) Microscopical illustration of micro-NMR of hyperpolarized fumarate in solution with NV ensembles in diamond. An ensemble of electronic NV center spins in diamond are optically initialized and readout. The NMR signal of fumarate in solution is detected by measuring the dipolar spin interaction of the electronic spins in diamond with nuclear spins in solution placed on top of the diamond surface. The external bias field $B_0$ induced by a surrounding electromagnet is chosen parallel to the NV quantization axis and tuned to about \qty{0.11}{\tesla}.
(b) Schematic view of the J-coupling spectroscopy protocol with varying echo intensity. Solid orange lines depict the magnetization in the horizontal plane of the Bloch sphere with the dashed line indicating the change of individual signal phases. QDyne detection sequences in between the CPMG pulses on the nuclear spins are tuned for detection of the proton signal, whose phase evolution is governed by the scalar coupling.
}
\label{fig:DNNR_protocol}
\end{figure*}

Micro- and Nanoscale NMR spectroscopy experiments have experienced a boost due to the use of NV centers in diamond for NMR detection in recent years~\cite{Schmitt2017a,Boss2017a,Glenn2018a,Staudenmaier2023a}.
However, even if combined with hyperpolarized substrate the technology currently operates for physiologically relevant concentrations at low SNR~\cite{Schwartz2019a,Bucher2020a,Arunkumar2021a,Smits2019a} and sample quantification alongside uncertainty estimation is a delicate issue.
Upon further improvement of the sensitivity, NMR detection with NV centers exhibits the potential for observation of metabolism on a single-cell level~\cite{Neuling2023a}.
We perform a microscale J-coupling spectroscopy experiment of hyperpolarized \CFum{}~\cite{Gierse2023a} using NV centers in diamond based on CPMG-like pulse scheme and detection of the Larmor precession during the echo periods.
The recorded data exhibits a two-dimensional structure, again, and is used for a further comparison of the AUC to the HML approach in the analysis.
\par

The measurement protocol is composed of CPMG pulses on \Cthirteen{} and \Hone{} spins, and thereby effectively reverts the free evolution of individual spin species, leaving a net evolution induced by the heteronuclear J-coupling.
Thus, the pulse protocol is specifically designed to differentiate molecules by the heteronuclear J-coupling which may serve as molecular fingerprint for the differentiation of metabolites.
When detecting the protons' Larmor precession signal during the echo, the effective evolution will manifest in a slow phase variation of the fast oscillating signal.
Assuming the phase to vary sufficiently slowly so that we can assume it constant over individual detection windows, we arrive at a description in which only the initial phase varies and, thereby, retrieve a scenario which is accessible by the analysis schemes presented in this work.
Figure~\ref{fig:DNNR_protocol}(a) schematically shows the detection setup~\cite{Striegler2025a} with further details given in \ref{app:nv_exp}.
Hyperpolarized \CFum{} is transferred from a reservoir to the diamond surface by a microfluidic channel, providing quick and reproducible sample placement.
\num{120} echoes are recorded over a period of \qty{20}{\milli\second} each, including the pulse and dead time amounting to approximately \qty{2.5}{\second}.
The short detection periods limit the Nyquist frequency to approximately \qty{50}{\Hz}, such that we only observe a single resonance in the direct dimension.
Plots showing the spectral data of each detection period and a 2D FFT plot are provided in figs.~S10-S11 of the supplementary material.
By renewal via the microfluidic channel, the experiment is repeated multiple times with a run-to-run separation of approximately \qty{7}{\second} using sample material from the same hyperpolarization batch.
\par

\begin{figure}[!bt]
\centering
\includegraphics[scale=1]{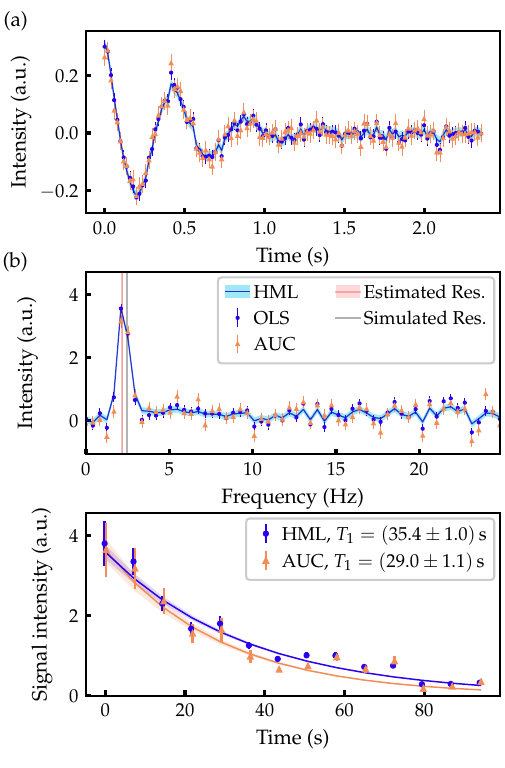}
\caption{
(a) Real part of the J-coupling induced evolution over time obtained via HML, OLS, and AUC estimates from NV detected micronscale experiments with hyperpolarized \CFum{}.
(b) Real part the FFT of (a). Compared to the AUC approach, the SNR is improved by \num{2.6} by the HML analysis. (c) The J-coupling signal intensity for different runs of the same experiment as in (a) upon renewal of the sample subject to $T_1$ relaxation.}
\label{fig:nv_data_plot}
\end{figure}

Given that we observe oscillations in both dimensions, a rather natural approach for the analysis is the two-dimensional FFT (see figs.~S10-S11 of the supplementary material).
This approach immediately provides insight into all the frequencies observed in the echo amplitude by providing us with a two-dimensional spectrum but it is not immediately clear how to perform a quantitative analysis from here and the exact procedure, moreover, might depend on the quantities of interest.
Instead, we will proceed to apply the analysis schemes discussed in sec.~\ref{sec:reaction_rate_est} and compare the outcomes.
To this end, we consider intensity estimates of individual echoes and the estimation of the overall signal intensity which is linked to the sample concentration and polarization.
The detection setup with NV centers records the sample magnetization in one direction only, which can, e.g., be obtained using only the real value of $\mvec{f}_T$ described in sec.~\ref{sec:reaction_rate_est}.
For better understanding, we choose a basis of $\sin$ and $\cos$ with the same frequency and twice as many but real valued $a_j$ to obtain an equivalent description.
Then, for each echo we are left with two amplitudes for $\sin$ and $\cos$, respectively, which we will refer to as the imaginary and real part.
The second level model describes the amplitude evolution imposed by the heteronuclear coupling.
The oscillation of real and imaginary part are described using the same frequency but individual amplitudes, again, employing a basis of trigonometric functions.
Figure~\ref{fig:nv_data_plot}(a) shows the real amplitude's evolution for each echo over time with estimates from the three different methods.
The corresponding FFT shown in fig.~\ref{fig:nv_data_plot}(b) indicates an SNR for the J-evolution of approximately \num{26} as compared to \num{13} for the VarPro and about \num{10} for the AUC approach.
Taking into account the finite precision and varying experimental conditions, the indicated frequency estimates from the HML approach and the resonance obtained from a simulation as described in \ref{app:nv_hamiltonian} using the couplings found in the literature~\cite{Eills2019a,Eills2023a} agree acceptably.
\par

Figure~\ref{fig:nv_data_plot}(c) shows the J-evolution's initial signal amplitude as shown in fig.~\ref{fig:nv_data_plot}(a)
for different consecutive runs of the experiment upon renewal of the sample via the microfluidic channel.
Because the basis functions $\sin$ and $\cos$ are practically orthogonal, the parameter estimation in a two-stage procedure almost perfectly coincides with the HML method and is, therefore, not shown here.
The observed decay is attributed to $T_1$ relaxation that occurs in the sample reservoir.
Even though the uncertainty on the $T_1$ estimates is almost identical, the difference in error bars hints at the HML method being more precise which is potentially relevant for more elaborate models.

\section{Summary and Discussion}
We investigated the use of a Bayesian hierarchical model for parameter estimation in two-predictor scenario assuming AWGN.
An analytic simplification led to an LS objective function, making it accessible to optimization using a number of well known routines, and providing gateway to uncertainty estimation from known principles.
In a simulation study for the metabolic conversion of hyperpolarized \CPyr{} to \CLac{} we demonstrated agreement of the result with the CRB and observed a reduction in uncertainty compared to intensity estimation by integration.
The analysis of experimental data recorded with a conventional NMR system, the method achieves uncertainties smaller by approximately a factor of \num{5} due to precise uncertainty propagation.
In a micronscale NMR experiment with NV centers, we have demonstrated the efficacy of the presented approach for an analysis of a two-dimensional NMR protocol and observed an effective improvement in SNR by a factor of 2 for the visualization of the evolution by heteronuclear scalar coupling for a sample of \CFum{}.
The observed signal resonance agrees with expectations based on existing literature within reasonable bounds.
\par

Hierarchical linear models with two predictors are frequently used in social sciences for several years, however, researchers typically aim to improve individual estimates by combination of different data sources~\cite{Raudenbush2002a,Woltman2012a}.
While marginalization is a well established procedure in Bayesian inference~\cite{Bretthorst1988a,Raimundez2023a}, it is rarely applied to hierarchical models, where MCMC sampling is used for numeric marginalization, instead.
We have applied marginalization over linear model parameters to a nonlinear hierarchical model and derived a novel procedure to parameter estimation in two-dimensional data scenarios.
The analytic elimination of parameters is computationally more efficient compared to numerical marginalization after MCMC sampling and, furthermore, enables analytic discussions of the expected estimates and propagated uncertainties in selected scenarios.
\par

Compared to multi-stage approaches for the analysis, the new method reduces the chance of variations and errors introduced by the user and, thus, delivers reliable and reproducible results and accurate uncertainty estimates.
Due to its accessibility through existing LS solving routines, the method can easily be integrated in existing frameworks for LS parameter estimation.
In the scenarios considered here, the spectral data are of high quality and can be reduced to a single resonance per compound by appropriate cropping of the time-domain data.
While the proposed approach can, in principle, also be applied to more complex experimental situations, this would require either a precise decomposition of the model into one basis function per compound or a generalization of the kinetic model to account for multiple resonances per compound, both of which would substantially increase the model complexity.
A major limitation of the HML approach is the assumption of invariant basis functions in the indirect dimension, which is hard to correct for.
Frequency drifts and changes in relative signal amplitudes for one compound, e.g., through different $T_1$ of different nuclei, might be more readily addressed in the AUC-based approach.
For data scenarios as observed in the NV data, on the other hand, the results clearly demonstrate that exploiting the full data set in a combined analysis helps to stabilize the estimation procedure and renders it robust with respect to colored noise.
Besides the two examples shown in this work, the method might be useful for the quantitative analysis of various kinds of NMR experiments involving periodic pulse sequences, such as CPMG experiments for the determination of $T_2$~\cite{Meiboom1958a}.
Beyond NMR, similar estimation problems are encountered in reaction rate estimation in time-resolved photospectroscopy~\cite{Bijlsma2000a,Bijlsma2002a,vanStokkum2004a}.

\section*{Acknowledgements}
L.~H.~B. thanks Koenraad Audenaert and Jan~F.~Haase for useful discussions.
P.~R.~J. and L.~H.~B. thank Magnus Karlsson for assistance with acquisition and analysis of the dDNP data used in Section~\ref{sec:reaction_rate_est}.
dDNP-NMR data were acquired with equipment funded by the Novo Nordisk Foundation (NNF 19OC0055825).
This work was supported by the BMBF through grant no 03ZU1110FF (QMED) and no 13N16447 (QuE-MRT).
The European Research Council provided support through the ERC Synergy Grant HyperQ (Grant No. 856432) and project C-QuENS (Grant No. 101135359).

\section*{Supplementary Material}
Supplementary material to this article is available at the end of this document.

\section*{Data Availability}
The supporting data and scripts used for processing are available at \url{https://doi.org/10.5281/zenodo.18338291}.

\bibliographystyle{elsarticle-num}

\appendix
\section{Derivation of the Hierarchical Model} \label{app:hierarchical_model}

\subsection{Motivation} \label{app:hm_motivation}
To motivate the specific hierarchical model used, consider data samples of the lower level model recorded for fixed $T$.
For better readability, the subscript that is used in the main part is dropped in the following discussion.
The standard expression for the likelihood distribution under assumption of white Gaussian noise reads
\begin{align}
\condprob{\mvec{y}}{\mvec{a},\omega} = \frac{1}{\sqrt{2\pi\sigma^2}^n}\exp \Big( \frac{1}{2\sigma^2}\Vert \mvec{y} - \Phi\mvec{a} \Vert^2 \Big) \label{eq:basic_likelihood}
\end{align}
with either $\mvec{y} \in \mathbb{R}^n$ or $\mvec{y} \in \mathbb{C}^n$ and parameters $\omega$ implicitly contained in $\Phi$.
$\sigma$ denotes the noise standard deviation which will be substituted by the precision $\beta_y = 1/\sigma^2$ from now on.
In the scenario of perfectly determined $\omega$, the matrix $\Phi$ is invariate and the estimation of amplitudes $\vec{a}$ from this likelihood distribution reduces to ordinary least-squares (OLS) estimation.
The corresponding estimator and covariance matrix are given by
\begin{align}
\est\mvec{a} = \Phi^\dagger \mvec{y} \ \text{ and } \ \Sigma_{\mvec{a}} = \frac{1}{\beta_y} (\transpose\Phi{} \Phi)^{-1} \ecomma \label{eq:OLScov}
\end{align}
respectively~\cite{Raudenbush2002a}.
Formally, the estimates are normally distributed random variables, which give rise to the distribution of $\mvec{a}$ as
\begin{align}
\begin{split}
\condprob{\mvec{a}}{\mvec{\mu}_a} = &\frac{1}{\sqrt{2\pi}^m} \sqrt{\det(\mmat{B}_a)} \\
&\times \exp(-\frac{1}{2}\transpose(\mvec{\mu}_a-\mvec{a}) \mmat{B}_a(\mvec{\mu}_a-\mvec{a})) \fullstop
\end{split}
\end{align}
Typically, as these are unbiased estimators, we expect them to vary around the amplitude's true value $\true\mvec{a} $, such that $\mvec{\mu}_a=\true\mvec{a} $.
For set up of the hierarchical model, however, the only information available about the expected value originates from the second level model.
Thus, upon substitution of $\mvec{\mu}_a$ by the corresponding amplitude value computed from the second level model, this distribution serves as basis for the prior distribution on $\mvec{a}$.
The inverse covariance matrix $\mmat{B}_a$ is substituted by the inverse of the covariance matrix in eq.~\eqref{eq:OLScov}, fixing the last missing ingredient to complete the model.
After some reordering, the prior distribution reads
\begin{align}
\begin{split}
\condprob{\mvec{a}}{\lambda} = &\sqrt{\frac{\beta_y}{2\pi}}^m \sqrt{\det(\transpose\Phi{} \Phi)} \\
	&\times \exp(-\frac{\beta_y}{2}\Vert \Phi \mvec{a}(\lambda) - \Phi \mvec{a} \Vert^2) \fullstop \label{eq:hierarchical_prior}
\end{split}
\end{align}
Above approach constitutes a combination of knowledge from the lower level model and the upper level model into a unified hierarchical picture.

\subsection{Analytic Marginalization Theorem} \label{app:marginalization}
Before discussion of the analytic marginalization of the posterior distribution in eq.~\eqref{eq:hierarchical_posterior}, we prove that
\begin{align}
\begin{split}
&\int_{\mathbb{R}^m} \mathrm{d}\mvec{a} \sqrt{\frac{\beta_y}{2\pi}}^{n} \exp(-\frac{\beta_y}{2} \Vert \mvec{y} - \Phi \mvec{a} \Vert^2) \\
&= \sqrt{\frac{\beta_y}{2\pi}}^{n-m} \frac{1}{\sqrt{\det (\transpose\Phi{} \Phi )}} \exp(-\frac{\beta_y}{2} \Vert \mvec{y} - \Phi\Phi^\dagger \mvec{y} \Vert^2 )
\end{split}
\label{eq:integration_corollary}
\end{align}
with $\Phi \in \mathbb{R}^{n\times m}, n\gg m, \rank(\Phi) = m$ and $\Phi^\dagger$ denotes the Moore-Penrose pseudoinverse.
We start by rewriting $\Phi$ with the QR decomposition into a product of a unitary and a reduced matrix
\begin{align}
\Phi = \mmat{U}\mmat{Q} \ \text{ with } \ \mmat{Q} = \begin{pmatrix}
\mmat{\tilde{Q}} \\
0
\end{pmatrix} \ecomma
\end{align}
with unitary matrix $\mmat{U}$ and $\mmat{\tilde{Q}}$ is $m\times m$ and invertible.
Proceed to rewrite the exponential argument under utilization of the invariance of the norm under unitary transformation as
\begin{align}
\begin{aligned}
\Vert \mvec{y} - \Phi\mvec{a} \Vert^2 &= \Vert \transpose\mmat{U} (\mvec{y} - \Phi\mvec{a}) \Vert^2 = \Vert \transpose\mmat{U} \mvec{y} - \mmat{Q}\mvec{a} \Vert^2 \\
&= \Vert (\idmat{}_m \oplus 0) (\transpose\mmat{U} \mvec{y} - \mmat{Q}\mvec{a}) \Vert^2 \\
	&\qquad + \Vert (0 \oplus \idmat{}_{n-m}) (\transpose\mmat{U} \mvec{y} - \mmat{Q}\mvec{a}) \Vert^2 \\
&= \Vert (\idmat{}_m \oplus 0) (\transpose\mmat{U} \mvec{y} - \mmat{\tilde{Q}}\mvec{a}) \Vert^2 \\
	&\qquad + \Vert (0 \oplus \idmat{}_{n-m}) \transpose\mmat{U} \mvec{y} \Vert^2 \fullstop \label{eq:exp1}
\end{aligned}
\end{align}
Note that the second term of the RHS of eq.~\eqref{eq:exp1} is completely independent of $\mvec{a}$ and can be rewritten under insertion of $\mmat{U}$ as
\begin{align}
\begin{aligned}
\Vert \mmat{U} (0 \oplus \idmat{}_{n-m}) \transpose\mmat{U} \mvec{y} \Vert^2 &= \Vert \mvec{y} - \mmat{U} (\idmat{}_m \oplus 0) \transpose\mmat{U} \mvec{y} \Vert^2 \\
&= \Vert \mvec{y} - \mmat{P}_\Phi \mvec{y} \Vert^2 \ecomma
\end{aligned}
\end{align}
where we introduced the projector $\mmat{P}_\Phi$.
Plugging above expression and the first term of eq.~\eqref{eq:exp1} back into the exponential argument of the LHS of eq.~\eqref{eq:integration_corollary} yields
\begin{align}
\begin{split}
&\sqrt{\frac{\beta_y}{2\pi}}^{n} \exp(-\frac{\beta_y}{2}\Vert \mvec{y} - \mmat{P}_\Phi \mvec{y} \Vert^2) \\
&\times  \int\limits_{\mathbb{R}^m} \mathrm{d}\mvec{a} \exp(-\frac{\beta_y}{2} \Vert \mmat{\tilde{Q}} \big( \mmat{\tilde{Q}}^{-1} (\idmat{}_m \oplus 0) \transpose\mmat{U} \mvec{y} - \mvec{a} \big) \Vert^2) \fullstop
\end{split}  \label{eq:corollary1_lhs}
\end{align}
The integral term constitutes a Gaussian integral and yields the normalization factor $\sqrt{(2\pi)^m \det(\Sigma)}$ with
\begin{align}
\det(\Sigma) =\frac{1}{[\beta_y \det(\transpose\mmat{\tilde{Q}} \mmat{\tilde{Q}})]} = \frac{1}{\beta_y \det(\transpose\Phi{} \Phi)]} \fullstop
\end{align}
Plugging the integration result back into eq.~\eqref{eq:corollary1_lhs} yields the RHS of eq.~\eqref{eq:integration_corollary}.
We now further note that the projector $\mmat{P}_\Phi$ can be expressed as
\begin{align}
\begin{aligned}
\mmat{P}_\Phi &= \mmat{U}(\idmat \oplus 0)\transpose\mmat{U} \\
	&= \mmat{U}(\mmat{Q} (\transpose\mmat{Q} \mmat{Q})^{-1} \transpose\mmat{Q} ) \transpose\mmat{U} \\
	&= \Phi (\transpose{\Phi} \Phi)^{-1} \transpose\Phi{} = \Phi\Phi^\dagger \ecomma
\end{aligned}
\end{align}
to complete the proof.
\par

The integrand in LHS of eq.~\eqref{eq:integration_corollary} represents the likelihood distribution of a simple estimation problem in $\omega$ and $\mvec{a}$ with one-dimensional data.
To find the distribution's maximum, we can equivalently minimize the negative logarithm
\begin{align}
\frac{\beta_y}{2} \Vert \mvec{y} - \Phi \mvec{a} \Vert^2 \fullstop
\end{align}
Golub and Pereyra~\cite{Golub1973a} have demonstrated, that this problem separates in the parameters $\omega$ and $\mvec{a}$, such that estimates of $\omega$ can be
derived from
\begin{align}
\frac{\beta_y}{2} \Vert \mvec{y} - \Phi\Phi^\dagger \mvec{y} \Vert^2 \ecomma \label{eq:reduced_LS_obj}
\end{align}
instead.
They obtain above expression by computing an analytic estimator for $\mvec{a}$ for fixed $\omega$ and plug it back into the original objective function.
Here, we have demonstrated that one obtains the same expression by marginalization and, thereby, decorate the reduced objective function \eqref{eq:reduced_LS_obj} with a statistical interpretation:
The reduced objective function corresponds to the marginal likelihood's logarithm and, thus, upon multiplication with a prior distribution, describes the posterior probability to observe parameter values $\omega$ given the data $\mvec{y}$.
Consequently, upon employing an accurate value for $\beta_y$, the width of this distribution, often approximated by the curvature of the exponential, gives rise to uncertainty estimates for $\omega$.
Point and uncertainty estimates of the amplitudes are then obtained from the maximum and width of $\condprob{\mvec{a}}{\est\omega{} , \mvec{y}}$, which for sufficiently flat prior distributions reduces to the known linear LS result.
Some other works suggest further marginalization of $\sigma$ upon utilization of Jeffrey's prior and end up with Student's t-distribution in place of the normal distribution~\cite{Bretthorst1988a,Bretthorst1990b}.
\par

In some early precision studies using the VarPro objective function, uncertainty estimates were based on repeated computation of the estimators for different realizations of noise and employing the Cramér-Rao bound~\cite{Veen1988a,Johnson1988a}.
Further works in the literature have employed analytic expressions of the Jacobian matrix for the VarPro formulation~\cite{Golub1973a,Kaufman1975a} and derived analytical expressions for Fisher information matrix assuming AWGN, but not providing an expression of the probability density after elimination of the linear parameters~\cite{Mullen2007a}.
By above note we have provided the missing link in this work and, by construction, recovered the efficiency of a maximum likelihood estimator which allows for saturation of the Cramér-Rao bound.
The ultimate covariance matrix can either be approximately computed from the Jacobian under assumption of vanishing residuals, potentially falling back to analytical expressions~\cite{Mullen2007a}, or by computing the inverse of the Hessian obtained from finite-differences.

\subsection{Hierarchical Marginal Likelihood} \label{app:HML}
We will now use the theorem in eq.~\eqref{eq:integration_corollary} to marginalize the product of likelihood~\eqref{eq:basic_likelihood} and hierarchical prior~\eqref{eq:hierarchical_prior}.
First, we rewrite the product's exponential argument
\begin{align}
\Vert \mvec{y} - \Phi \mvec{a} \Vert^2 + \Vert \Phi \mvec{a}(\lambda) - \Phi \mvec{a} \Vert^2 \fullstop \label{eq:exp_product}
\end{align}
by introducing
\begin{align}
\tilde{\mvec{y}} = \frac{1}{2} (\mvec{y} + \Phi \mvec{a}(\lambda)) \label{eq:ytilde}
\end{align}
to
\begin{align}
\frac{1}{2} \Vert \mvec{y} - \Phi \mvec{a}(\lambda) \Vert^2 + 2\Vert \tilde{\mvec{y}} - \Phi \mvec{a} \Vert^2 \fullstop \label{eq:posterior_log_rewritten}
\end{align}
We note that the first term of above expression is completely independent of $\mvec{a}$ and, thus, write it in front of the integral.
The second term is proportional the exponential argument of the LHS of eq.~\eqref{eq:integration_corollary} and we can immediately write down the integration result of the exponential of the second term as
\begin{align}
\sqrt{\frac{\pi}{\beta_y}} \frac{1}{\sqrt{\det(\transpose\Phi{} \Phi)}} \exp(-\beta_y \Vert \tilde{\mvec{y}} - \mmat{P}_\Phi \tilde{\mvec{y}} \Vert^2 ) \fullstop
\end{align}
Substitution of eq.~\eqref{eq:ytilde} in the exponential yields
\begin{align}
\frac{1}{4} \Vert \mvec{y} - \mmat{P}_\Phi \mvec{y} \Vert^2
\end{align}
for the exponential argument.
Upon inserting this result into the posterior expression, the normalization factor $\sqrt{\det(\transpose\Phi{} \Phi)}$ is canceled out and we are left with
\begin{align}
\begin{aligned}
\int_\mathbb{R} &\mathrm{d}\mvec{a} \condprob{\mvec{y}}{\mvec{a},\omega}\condprob{\mvec{a}}{\lambda} = \left(\frac{\beta_y}{2\pi}\right)^{n/2} \cdot \left(\frac{1}{2}\right) ^{\frac{m}{2}} \\
&\times \exp\lbrace-\frac{\beta_y}{2} \frac{1}{2}[ \Vert \mvec{y}-\Phi\mvec{a}(\lambda) \Vert^2 + \Vert\mvec{y} - \Phi\Phi^\dagger \mvec{y} \Vert^2 ] \rbrace \fullstop
\end{aligned}
\end{align}
The choice of the distributions implies independent noise for different $T$, such that above expression trivially generalizes to multiple values of $T$ by muliplication of individual probabilities.
For data recorded at $k$ different $T$, the marignal likelihood then reads
\begin{align}
\begin{aligned}
&\mathcal{L}(\omega,\lambda) := \left(\frac{\beta_y}{2\pi}\right)^{\frac{nk}{2}} \cdot \left(\frac{1}{2}\right) ^{\frac{m}{2}} \\
&\times \exp\lbrace-\frac{\beta_y}{2} \frac{1}{2}( \Vert \mmat{Y}-\Phi\mmat{A}(\lambda) \Vert^2 + \Vert\mmat{Y} - \mmat{P}_\Phi \mmat{Y} \Vert^2 ) \rbrace \ecomma
\end{aligned}
\end{align}
where matrix $\mmat{Y}$ is composed of column vectors of data and $\mmat{A}(\lambda)$ of columns vectors of $\mvec{a}_T(\lambda)$ for different $T$.
The logarithm of above expression then constitutes the objective function stated in eq.~\eqref{eq:neglogl} of the main text.

\subsection{Amplitude Estimates} \label{app:ampl_est}
The Bayesian recipe for hierarchical models allows us to derive estimates for amplitudes $\vec{a}$ from the maximum of
\begin{align}
\condprob{\mvec{y}}{\mvec{a},\est\omega}\condprob{\mvec{a}}{\est\lambda} \varpropto \exp \Big(-\frac{\beta_y}{2}\cdot 2\Vert \tilde{\mvec{y}} - \Phi \mvec{a} \Vert^2_2 \Big) \ecomma
\end{align}
where we used the exponential of expression~\eqref{eq:posterior_log_rewritten} and dropped all terms not depending on $\mvec{a}$.
The values of $\est\omega{} $ and $\est\lambda{} $ are implicitly contained in $\tilde{\mvec{y}}$ and $\Phi$.
The exponential argument of above expression is known from the OLS problem and, thus, under substitution of $\tilde{\mvec{y}}$ as in eq.~\eqref{eq:ytilde}, we obtain expression~\eqref{eq:hierarchical_ampl_estimate}.
The covariance matrix is likewise constructed from $\Phi$ under consideration of the additional factor of $2$ in front, yielding expression~\eqref{eq:hierarchical_ampl_covariance}, which is exactly smaller by a factor of $1/2$ compared to the OLS covariance.
This manifests in a rescaling by $1/\sqrt{2}$ of the individual uncertainty estimates.

\subsection{Reformulation in the Standard LS Form} \label{app:LS_form}
The standard form of LS problems reads
\begin{align}
\min\limits_x \frac{1}{2\sigma^2} \Vert \mvec{r}(x) \Vert^2 \ecomma
\end{align}
with $x\in \mathbb{R}^l$, $r: \mathbb{R}^l \rightarrow \mathbb{R}^k, x\mapsto \mvec{r}(x)$.
To rewrite the objective function in expression~\eqref{eq:neglogl} in the corresponding form, focus on the case of fixed $T$ first.
Then
\begin{align}
\begin{aligned}
\frac{\beta_y}{2} \frac{1}{2} ( \Vert \mvec{y} - \mmat{P}_\Phi \mvec{y} \Vert^2 + \Vert \mvec{y} - \Phi \mvec{a}(\lambda) \Vert^2 ) \\
 = \frac{\beta_y}{2} \Biggl\Vert \underbrace{\frac{1}{\sqrt{2}}
\begin{pmatrix}
\mvec{y} - \mmat{P}_\Phi \mvec{y} \\
\mvec{y} - \Phi \mvec{a}(\lambda)
\end{pmatrix}}_{=:\mvec{r}(\omega,\lambda)}
\Biggr\Vert^2 \fullstop
\end{aligned}
\end{align}
For the generalization of $\mvec{r}(\omega,\lambda)$ to data recorded at multiple $T$, the most straightforward approach is to organize all matrix columns in a correspondingly larger vector.
The use of complex-valued signals extends this vector's length by an additional factor of two.
In an implementation, this operation is sufficiently achieved by providing a one-dimensional view to the matrix data.
\par

The Jacobian of $\mvec{r}$ assuming $\omega$ and $\lambda$ of size one, each, reads
\begin{align}
J_{\mvec{r}} = \frac{1}{\sqrt{2}}\begin{pmatrix}
\pardiff{\mmat{P}_\Phi^\perp }{\omega} \mvec{y} & \mvec{0} \\
\pardiff{\Phi}{\omega} \mvec{a}(\lambda) & \Phi\pardiff{\mvec{a}(\lambda)}{\lambda}
\end{pmatrix} \ecomma \label{eq:rjac}
\end{align}
where we have introduced the orthogonal projector $\mmat{P}_\Phi^\perp = \idmat - \mmat{P}_\Phi$.
A formula for the computation of its partial derivative is provided in ref.~\cite{Golub1973a}.
Extension of $\omega$ and $\lambda$ to larger tuples leads to insertion of additional columns with the corresponding partial derivative.
For an expression covering multiple $T$, the Jacobian also has to be adjusted accordingly.
An analytic evaluation of the Hessian matrix under assumption of vanishing residuals is possible using results presented in~\cite{Mullen2007a}.
The estimates computed in this work, however, solely rely on the numerical computation of the Jacobian, i.e., eq.~\eqref{eq:rjac} and correspondingly derived structures have not been used, because it vastly inflates the implementation complexity without guarantee for computational advantage.

\subsection{Notes on the Use of Polar Coordinates} \label{app:polar_coords}
Second-level models for reaction rate estimation are typically associated with signal intensity and, therefore, map to $\mathbb{R}^m$.
In the formulation presented in the main text, the second level model is required to span the complex domain if the basis function does so.
From a mathematical perspective, an extension of the model is trivially possible by introduction of additional phase parameters and multiplying the amplitudes with a complex phase factor (i).
Mathematically, for the experiment treated in Section~\ref{sec:reaction_rate_est}, this would permit us to use basis functions
\begin{align}
\exp\big( \iu \omega_j t \big)  \, ,
\end{align}
instead, and the phases to be estimated then only show up in the second sum term of eq~\eqref{eq:neglogl}.
Alternatively, we can stick to the approach described in Section~\ref{sec:reaction_rate_est} and introduce the same number of additional phases to include them in the first level model (ii).
For complex-valued data, this option corresponds to a description using polar coordinates with basis functions
\begin{align}
\exp\big[ \iu (\omega_j t + \varphi_j) \big]
\end{align}
and using real-valued amplitudes $\rho_j$.
For a compact notation for complex-valued signals, we can reorganize the basis matrix $\phi + \iu \theta$ and data $\mvec{u}+\iu \mvec{v}$ accordingly to
\begin{align}
\Phi = \begin{pmatrix}
\phi \\ \theta
\end{pmatrix}
\ \text{ and } \
\mvec{y} = \begin{pmatrix}
\mvec{u} \\ \mvec{v}
\end{pmatrix}
\end{align}
with $\phi$ and $\theta$, as well as $\mvec{u}$ and $\mvec{v}$ denoting the real and imaginary parts, respectively, as briefly described in the main text.
The phase now shows up in both terms of eq.~\eqref{eq:neglogl} and being stronger restricted by the data in the estimation procedure.
\par

Conceptually, the two approaches correspond to choosing prior distributions for either of cartesian or polar coordinates which we will marginalize over.
For option (i), we integrate over cartesian coordinates $\mathrm{d}a_r\mathrm{d}a_i$ and, thereby, implicitly marginalize over the signal phase, which we later have to add again in the second-level model.
When choosing option (ii), the marginalization over the phase is not trivially possible~\cite{Jaynes1987a} and one has to restrict to integration over $\mathrm{d}\rho$.
The total degrees of freedom, however, are invariate regardless of the option chosen.
Ultimately, the specific choice impacts the implications posed about the prior knowledge included into the model~\cite[Appendix D]{Bretthorst1988a}.
Because option (i) restricts the signal intensities to real values and , therefore, resembles our prior knowledge more precisely, we stick to the polar prior for the analysis of the pyruvate reaction data.
In numerical experiments using the same scenario as in sec.~\ref{sec:reaction_rate_est}, we have observed identical estimation results for the reaction rate estimate for both approaches, (i) and (ii), up to estimator variance.
The data, however, in addition to drifts in the resonance frequencies exhibits variation in the signal phase and the two approaches do not yield identical results.

When motivating the choice of the prior distribution purely from the prior knowledge and not paying special attention to the propagation of uncertainty up the hierarchy, we typically understand the kinetic model as a model for $\rho(T;\lambda)$.
The fact that $\rho$ represents an amplitude in the sense of $\rho=\vert a \vert$, the prior should reflect that $\rho \in [0,\infty)$, accordingly.
Moreover, different $\rho_j$ associated with different signal resonances are generally correlated, such that the prior distribution does not factorize, but constitutes a multivariate distribution instead with correlations being treated as hyperparameters.
A simple guess for the prior could be a truncated Gaussian distribution limited to the positive range.
Analytical marginalization (independent of the exact shape of the covariance or whether it is fixed or variable) does not yield a closed form expression that can easily be treated for a large set of FID runs.
An alternative reasonable candidate is the Rice distribution~\cite{Rice1944a,Rice1945a,Yakovleva2018a}, because it would appropriately reflect the link of $\rho = \vert a \vert$ whose real and imaginary part are independently normally distributed.
A generalization to a multivariate scenario including correlations, however, does not even yield a closed form expression for the probability distribution.
In that case a full treatment via MCMC is theoretically possible but will inevitable fail computationally for the data dimensions in this work.

\section{Experimental Details of the Metabolic Conversion and Data Processing} \label{app:hela_exp}
\subsubsection*{Culturing of HeLa cells}
The human cervical cancer cell line HeLa (American Type Culture Collection, ATCC CCL-2, Manassas, VA) was routinely maintained in Dulbecco's Modified Eagle Medium (DMEM; Biowest L0103) with \qty{4.5}{\gram\per\litre} D-glucose, stable L-glutamine, sodium pyruvate) with the addition of \qty{10}{\percent} fetal bovine serum (Biowest S1810), and \qty{1}{\percent} penicillin/streptomycin (Biowest L0022) at \qty{37}{\celsius} in an \qty{5}{\percent} CO2 atmosphere.
The cells were maintained by replacing the growth medium \numrange{2}{3} times per week.
HeLa cells were seeded at \numproduct{0.5 x 120} cells per T75 flask \num{3} days before the experiment day.
Confluent HeLa culture was harvested by trypsinization (\qty{0.25}{\percent} Trypsin-EDTA solution, Biowest X0930), and detached cells were centrifuged (\qty{200}{\rpm}, \qty{2}{\min}) and washed in DMEM with FBS.
Hereafter the cells were washed in PBS with \ce{Ca^{2+}/Mg^{2+}} (HyClone, SH30264.01) and centrifuged again.
Finally, the cells were resuspended to \qty[parse-numbers=false]{10 \times 106}{cells\per\milli\litre} in PBS with \ce{Ca^{2+}/Mg^{2+}} and \qty{200}{\micro\litre} were transferred to a Shigemi NMR tube.

\subsection*{dDNP-NMR Experiments}
A substrate sample stock solution was prepared from [1-\ce{^{13}C}] pyruvic acid (Sigma-Aldrich) doped with trityl radical AH111501 (GE Healthcare) to \qty{17}{\milli\Molar} and Gadoteridol gadolinium chelate solution (Bracco Imaging) to \qty{1.5}{\milli\Molar}.
For each experiment \qty{3.1}{\milli\gram} sample was hyperpolarized to equilibrium polarization in a Hypersense \qty{3.3}{\tesla} polarizer (Oxford Instruments).
After complete hyperpolarization buildup (approximately \qty{1}{\hour}) the sample was dissolved in \qty{5}{\milli\litre} phosphate buffer (pH \num{7.4}, \qty{40}{\milli\Molar}) with added \qty{3}{\micro\litre} of a \qty{10}{\Molar} \ce{NaOH} solution.
The dissolution parameters were set to produce a temperature of the solution which, after transport and injection into the cell suspension was approximately \qty{310}{\kelvin}.
The concentration of pyruvate in the solution, after dissolution, was \qty{7}{\milli\Molar}, and the liquid state polarization was approximately \qty{25}{\percent}.
After manual injecting of the hyperpolarized \CPyr{} solution into the cell suspension, a time series of \ce{^{13}C} NMR spectra was recorded using a pulse angle of \qty{10}{\degree} and a \qty{2}{\s} total delay between pulses.
The NMR data were recorded on a Bruker \qty{500}{\mega\Hz} AVANCE NEO spectrometer fitted with a \qty{5}{\mm} DCH cryoprobe.
Spectral plots of the raw data is shown in figs.~S1-S3 of the supplementary material.

\subsection*{Data Post-Processing}
Estimates derived from the AUC approach are obtained upon integration of the area under the curve using the Bruker software TopSpin.
Prior to integration, phase and baseline corrections are applied.
There are no error bars estimated but the variation is computed from the fluctuation of data points around the model.
\par

For time domain analysis the data is exported and the digital filter imposed by Bruker is removed using methods implemented in the Python package NMRglue~\cite{Helmus2013a}.
Supposedly because of initial changes of temperature and pH of the sample upon mixing with the hyperpolarized pyruvate, the resonances exhibit a slight drift over the first few FID runs.
Towards later $T$, the resonance frequencies appear to approach an equilibrium value.
In order to apply the theory introduced in this paper, about the Pyruvate resonances of the first \num{50} FID runs are fitted individually and the resonance frequencies are corrected in time domain by subtracting the signal fitted and adding it with a slightly corrected frequency.
Furthermore, to practically observing a single resonance line only, the data of individual FID runs is being cropped to about \qty{150}{\milli\second}, which roughly corresponds to two times the observed $T_2^\ast$.
Plots showing the data before and after drift correction and cropping are provided in figs.~S1-S3 along with plots showing the frequency change and influence on the amplitudes in figs.~S4-S6 of the supplementary material.
The noise parameter $\beta_y = 1/\sigma_y^2$ required for uncertainty estimation is estimated from the residuals at the optimal parameters.
The data exhibited \CPyr{} (hydrate) as additional major resonance which has to be accounted for in the time-domain analysis.
The kinetic model used for the HML approach is, therefore, extended by an additional resonance which is assumed not to contribute to the reaction but follows a simple exponential decay over time.
The optimization and least-squares fitting is performed using the SciPy~\cite{Virtanen2020a} package in Python.
\par

\subsection*{Simulation Parameters}
The simulation is performed using the models described in the main text combined with parameter values provided in table~\ref{tab:sim_pars} assuming unit time, i.e., sampling intervals equal \qty{1}{\s} in both dimensions.
For the statements derived from the simulation to also apply to real experiments, the parameter values used are rounded fit results of the third sample presented in fig.~\ref{fig:results_plot}~(b).
AWGN is simulated by addition of normally distributed random numbers to the signal model.
\begin{table}
\centering
\begin{tabular}{c|c||c|c}
\toprule
Par. & Value & Par. & Value\\
\midrule
$P_0$ & 9.756 & $\omega_\mathrm{P}$ & \qty{1.826}{\Hz} \\
$\kappa_\mathrm{P}$ & \qty{0.060}{\Hz} & $\kappa_{\omega,\mathrm{P}}$ & \qty{1.006e-3}{\Hz} \\
$L_0$ & 0.012 & $\varphi_\mathrm{P}$ & \num{0.0} \\
$\kappa_\mathrm{L}$ & \qty{0.013}{\Hz} & $\omega_\mathrm{L}$ & \qty{2.145}{\Hz} \\
$k$ & \qty{8.78e-4}{\Hz} & $\kappa_{\omega,\mathrm{L}}$ & \qty{1.302e-3}{\Hz} \\
& & $\varphi_\mathrm{L}$ & \num{0.0} \\
\bottomrule
\end{tabular}
\caption{Parameter values used for the simulation presented in sec.~\ref{sec:reaction_rate_est} in unit time scales, i.e., sample intervals are assumed \qty{1}{\s} in both dimensions. To recover the parameters in real units divide rates and frequencies by the sampling interval.}
\label{tab:sim_pars}
\end{table}

\section{Details of the micro-NMR Setup} \label{app:nv_exp}
\subsection*{NV Detection Setup}
The diamond sensor used has a flat surface with a densely packed layer of NV centers (2 ppm), averaging a depth of approximately 10 µm.
The initialization and readout of the electron spin is facilitated by 100 mW of \qty{532}{\nano\meter} laser light.
A custom-built optical microscope, integrated within an electromagnet that generates a magnetic field strength of around \qty{0.11}{\tesla}, provides optical spin access.
Precision in localization is achieved by focusing the laser spot to a diameter of about \qty{10}{\micro\meter}. The detector collects a large field of view to improve collection efficiency.
Microwave and radio frequency control are administered through a transmission line and a small millimeter-sized coil, positioned in proximity to both the sensor and the specimen spins.
~\cite{Striegler2025a}
\par

\subsection*{Sample Preparation}
To enable the injection of hyperpolarized specimens in solution, a microfluidic chip is integrated around the diamond NMR sensor.
Hyperpolarized fumarate solutions are prepared using an NVision Imaging Technologies GmbH setup, which employs parahydrogen-induced polarization.
The resulting sample has a concentration of approximately \qty{100}{\milli\Molar}, with initial nuclear hyperpolarization estimated to be around \qty{10}{\percent}.
~\cite{Gierse2023a}
\par

\subsection*{Detection Protocol and Data Post-Processing}
The detection protocol is characterized by a temporal interval of approximately \qty{19.8}{\milli\second} between CPMG pulses on both nuclear spin species.
For the QDyne protocol, the pulse distance is tuned to ensure sensitivity to the proton Larmor frequency.
During data post-processing, phase cycling corrections are conducted by applying a constant phase factor.
The optimization and least-squares fitting is performed using the SciPy~\cite{Virtanen2020a} package in Python.

\section{Theoretical Description of the J-Coupling Spectroscopy Protocol} \label{app:nv_hamiltonian}
The protocol used in section~\ref{sec:microNMR} aims to probe the J-coupling features of the considered molecule.
Typical dimensions of heteronuclear scalar couplings in a molecule are in the range of \qtyrange{1}{8}{\Hz}.
Therefore, the evolution is rather slow and to observe multiple oscillations, the signal must persist for longer than \qty{0.5}{\s}.
To overcome the limit of $T_2^\ast$ typically being smaller that \qty{0.5}{\s}, we employ a CPMG sequence on both species that refocuses the signal.
Starting from a simple Hamiltonian to describe the interaction between the probe spin $\mop{S}$ and the remaining protons $\mop{I}_j$ with $J=1,\dots, n$
\begin{align}
\begin{aligned}
\mop{H} &= \omega_S \mop{S}_z + \sum\limits_j \omega_j \mop{I}_z \\
 &+ \sum\limits_j J_j \mop{\mvec{S}} \cdot \mop{\mvec{I}}^j +  \sum\limits_{j<k} J_{jk} \mop{\mvec{I}}^j \cdot \mop{\mvec{I}}^k \\
 &+ \mop{H}_{\mathrm{RF},S} + \mop{H}_{\mathrm{RF},I}
\end{aligned}
\end{align}
In rotating frame w.r.t. to the control Hamiltonian's driving frequencies and focusing on the periods without driving field, the Hamiltonian simplifies under rotating wave approximation to
\begin{align}
\begin{aligned}
\mop{H}_\mathrm{int} &= \delta_S \mop{S}_z + \sum\limits_j \delta_j \mop{I}_z \\
 &+ \sum\limits_j J_j \mop{S}_z \mop{I}_z^j +  \sum\limits_{j<k} J_jk \mop{\mvec{I}}^j \cdot \mop{\mvec{I}}^k \label{eq:Hint}
\end{aligned}
\end{align}
Under a dynamical decoupling the disorder terms in the first line of eq.~\eqref{eq:Hint} accumulate to zero at the exact center between control pulses.
The terms in the second line only appear in quadratic order, such that they remain invariant under dynamical decoupling in an effective picture and the effective evolution is given by the coupling terms only, enabling J-coupling spectroscopy.
The emitted ac signal in the lab frame, however, still oscillates at the Larmor frequency of the nuclear spins and can be computed in x-direction, for example, by also transforming the $\mop{S}_x$ operator into the rotating frame to find
\begin{align}
\begin{aligned}
\expval{\mop{S}_x^{lab} (t)} = &\cos(\omega_{\mathrm{RF},S}) \expval{\mop{S}_x^{int} (t)} \\
 &+ \sin(\omega_{\mathrm{RF},S}) \expval{\mop{S}_y^{int} (t)}
\end{aligned}
\end{align}
Expectation values in the interaction frame can be computed numerically by evolving the system's state using the Hamiltonian in eq.~\eqref{eq:Hint} interrupted by CPMG pulses.

In the simplest approach, one now constructs a protocol that takes a single sample perfectly alternating between the CPMG pulses to also correct the fast oscillation and to directly observes the evolution by the J-coupling.
The detection protocol used in this work, however, records $n$ samples of the fast oscillating signal over a finite period between control pulses.
When choosing the pulse distance of CPMG sequences much shorter than the timescales of J-coupling dynamics, the expectation values of $\mop{S}_x,\mop{S}_y$ in the interaction frame can be assumed constant over individual sampling periods and the model amplitude central between two pulses encodes the dynamics we wish to observe.
Using $n$ samples from the oscillating signal per period immediately improves the estimates by a factor of $\sqrt{n/2}$.
\par

\titleformat{\section}
	[hang]
	{\large\bfseries}
	{S\arabic{section}.}
	{0.5em}
	{}
	[]
\renewcommand\thefigure{S.\arabic{figure}}
\renewcommand{\theequation}{S.\arabic{equation}}
\setcounter{section}{0}
\setcounter{figure}{0}

\newpage
\vspace*{50mm}
\begin{centering}
	\Huge{Supplementary Material to "\sharetitle"}
\end{centering}
\newpage

\section{Frequency Drift Correction in Time Domain}
\begin{figure}
	\includegraphics[scale=1]{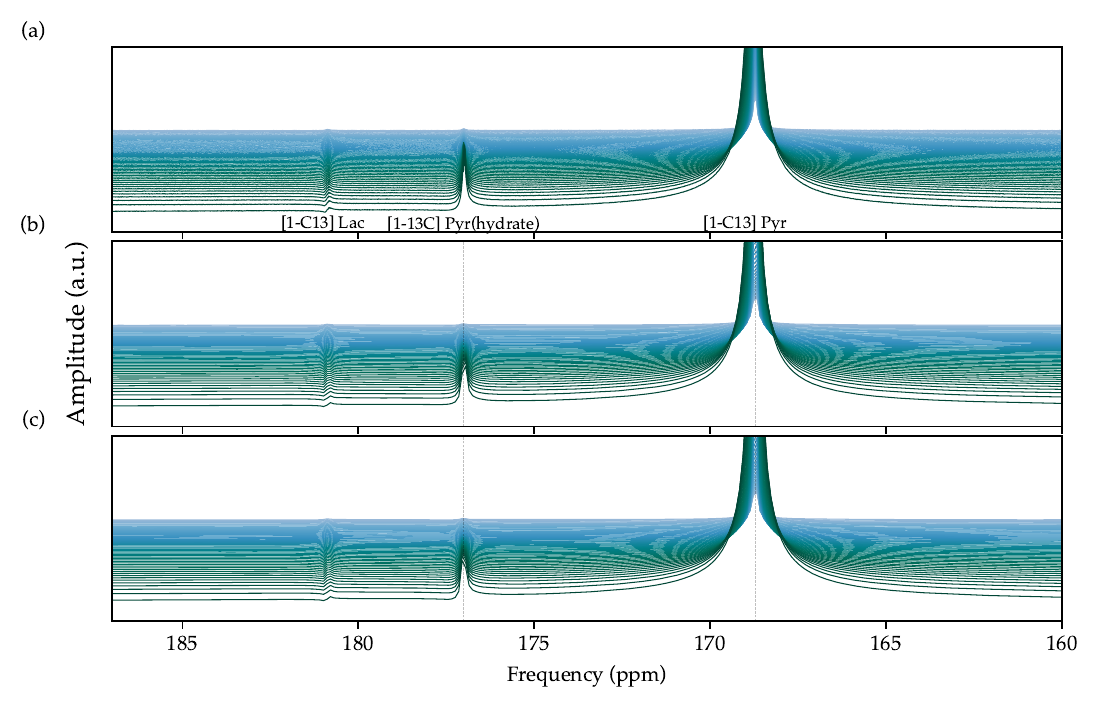}
	\caption{First experimental run: (a) The plot shows an excerpt of the absolute value of the data after FFT. (b) The same excerpt as in (a) after chopping the data after approx. \qty{150}{\milli\second}. Vertical dashed lines indicate the resonance frequency approached with increasing number of FID runs, which is also used for the correction. (c) The same excerpt after the correction of frequency drifts.}
	\label{fig:data_correction_2}
\end{figure}

\begin{figure}
	\includegraphics[scale=1]{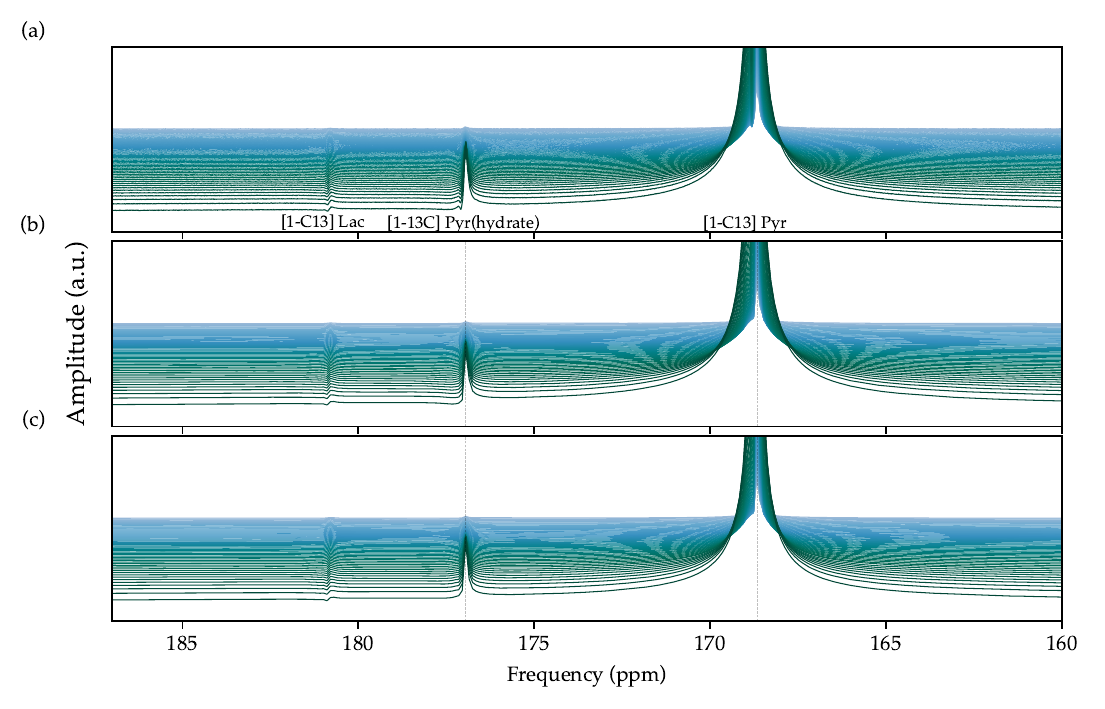}
	\caption{Second experimental run: (a) The plot shows an excerpt of the absolute value of the data after FFT. (b) The same excerpt as in (a) after chopping the data after approx. \qty{150}{\milli\second}. Vertical dashed lines indicate the resonance frequency approached with increasing number of FID runs, which is also used for the correction. (c) The same excerpt after the correction of frequency drifts.}
	\label{fig:data_correction_3}
\end{figure}

\begin{figure}
	\includegraphics[scale=1]{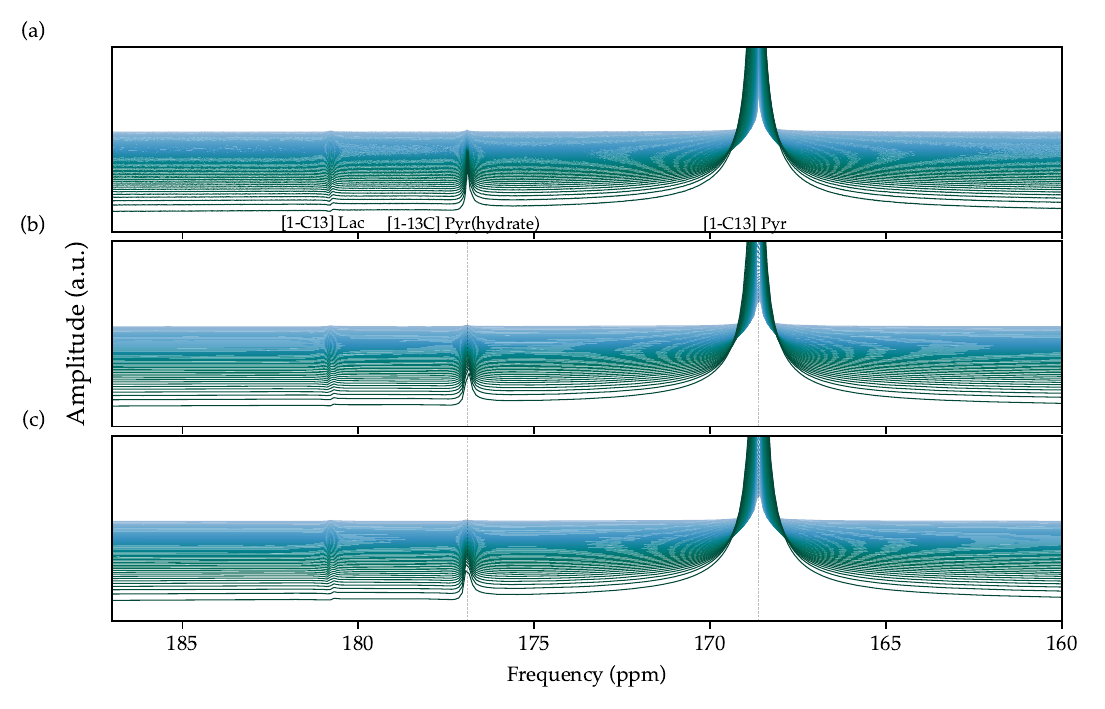}
	\caption{Third experimental run: (a) The plot shows an excerpt of the absolute value of the data after FFT. (b) The same excerpt as in (a) after chopping the data after approx. \qty{150}{\milli\second}. Vertical dashed lines indicate the resonance frequency approached with increasing number of FID runs, which is also used for the correction. (c) The same excerpt after the correction of frequency drifts.}
	\label{fig:data_correction_4}
\end{figure}

\begin{figure}
	\includegraphics[scale=1]{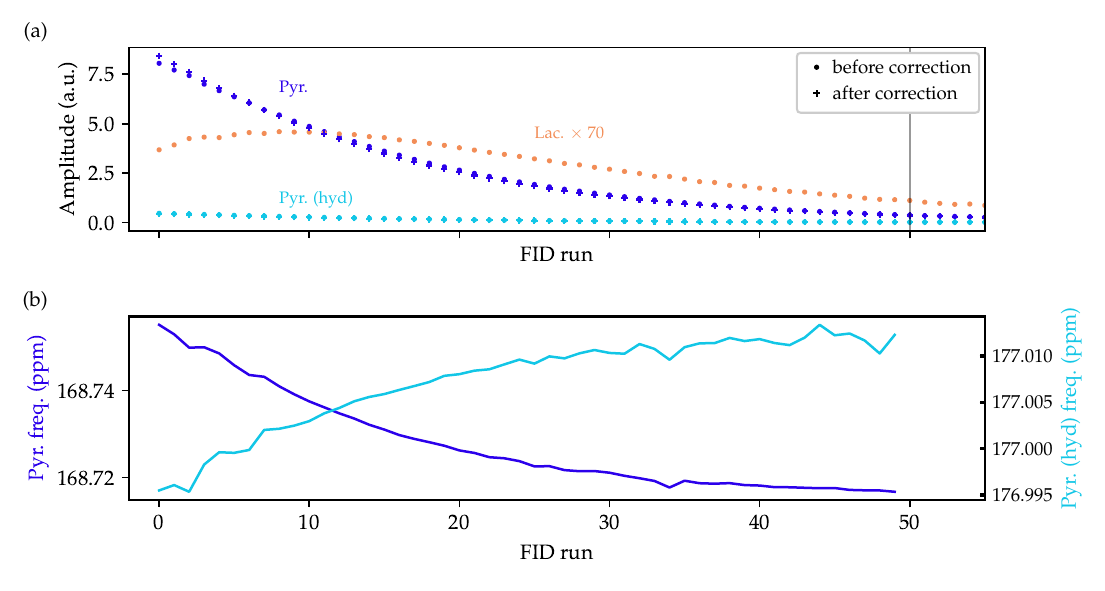}
	\caption{First experimental run: (a) Amplitude estimates from linear least-squares estimation before and after the drift correction. The Lactate amplitude is only shown before drift correction because this component is not corrected. The vertical gray line indicates the FID run from which the signal is corrected. (b) Evolution of the signal frequency over time for the two corrected resonances.}
	\label{fig:freq_drift_2}
\end{figure}

\begin{figure}
	\includegraphics[scale=1]{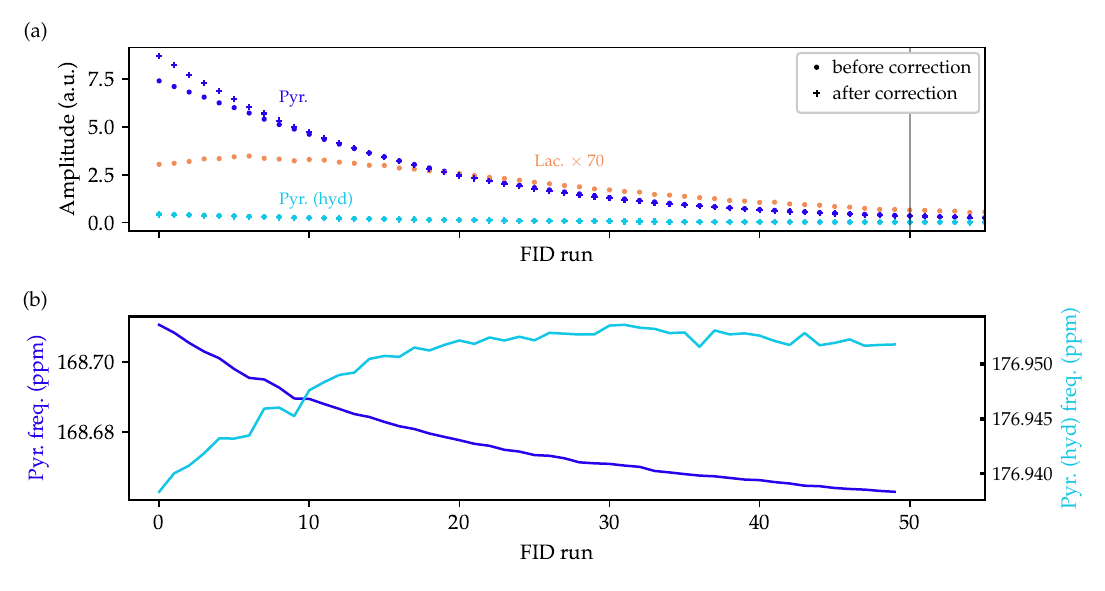}
	\caption{Second experimental run: (a) Amplitude estimates from linear least-squares estimation before and after the drift correction. The Lactate amplitude is only shown before drift correction because this component is not corrected. The vertical grey line indicates the FID run from which the signal is corrected. (b) Evolution of the signal frequency over time for the two corrected resonances.}
	\label{fig:freq_drift_3}
\end{figure}

\begin{figure}
	\includegraphics[scale=1]{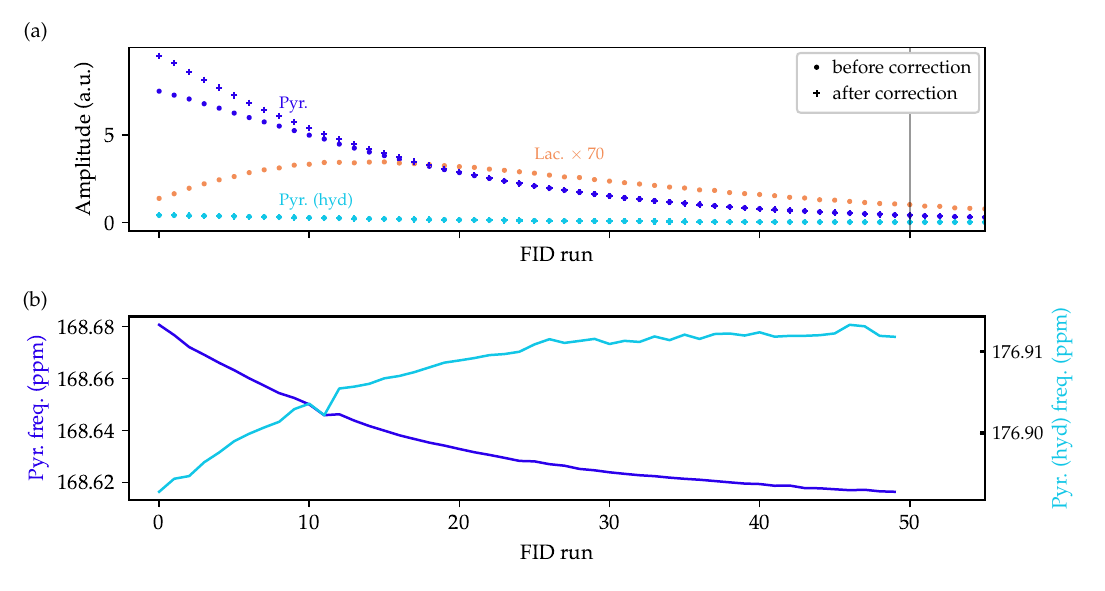}
	\caption{Third experimental run: (a) Amplitude estimates from linear least-squares estimation before and after the drift correction. The Lactate amplitude is only shown before drift correction because this component is not corrected. The vertical gray line indicates the FID run from which the signal is corrected. (b) Evolution of the signal frequency over time for the two corrected resonances.}
	\label{fig:freq_drift_4}
\end{figure}

In Appendix~B of the main text, we refer to a manual drift correction applied before performing the time-domain analysis.
The observed frequencies typically evolve between different FID runs and eventually approach a stable frequency point.
We use the stable frequencies approached in later runs of the experiment and correct the previous ones by fitting the signal in time domain, subtracting frequency components and adding them at the target frequency.
The steady frequency is approximated by the frequency determined from the 50th run and all previous FIDs are corrected sequentially.
Phase and amplitude of the signal remain completely untouched by this procedure.
Figures~\ref{fig:data_correction_2}, \ref{fig:data_correction_3}, and \ref{fig:data_correction_4} show the processing steps the data used in Section~3.1 of the main text undergoes graphically, corresponding to the data points in fig.~2(b) in the order of left-to-right.
Because there is no manual phase correction performed in these steps, the figure shows an excerpt of the absolute value of the data after FFT.
Plot (a) each is the raw data after removal of the digital filter using functionalities from nmrglue~\cite{Helmus2013a}.
Plot (b) each depicts the data after chopping after approx. \qty{150}{\milli\second}.
While before the chopping, potential splitting of the signals could have been identified, the signals collapse into a single resonance after the reduction of data.
Taking the dashed vertical lines as reference, the slight drift in frequency, however, is clearly visible.
The correction is only applied to the two major resonances coming from the Pyruvate because the drift in the Lactate resonance is small enough to be negligible.
Plot (c) each finally shows the spectrum after drift correction.
The evolution of the corrected frequencies are plotted over the run in figs.~\ref{fig:freq_drift_2}, \ref{fig:freq_drift_3}, and \ref{fig:freq_drift_4} in the bottom plot with the estimated amplitude values in the top plot each.

\section{Residuals of the Kinetic Model Fit}
\begin{figure}
	\includegraphics[scale=1]{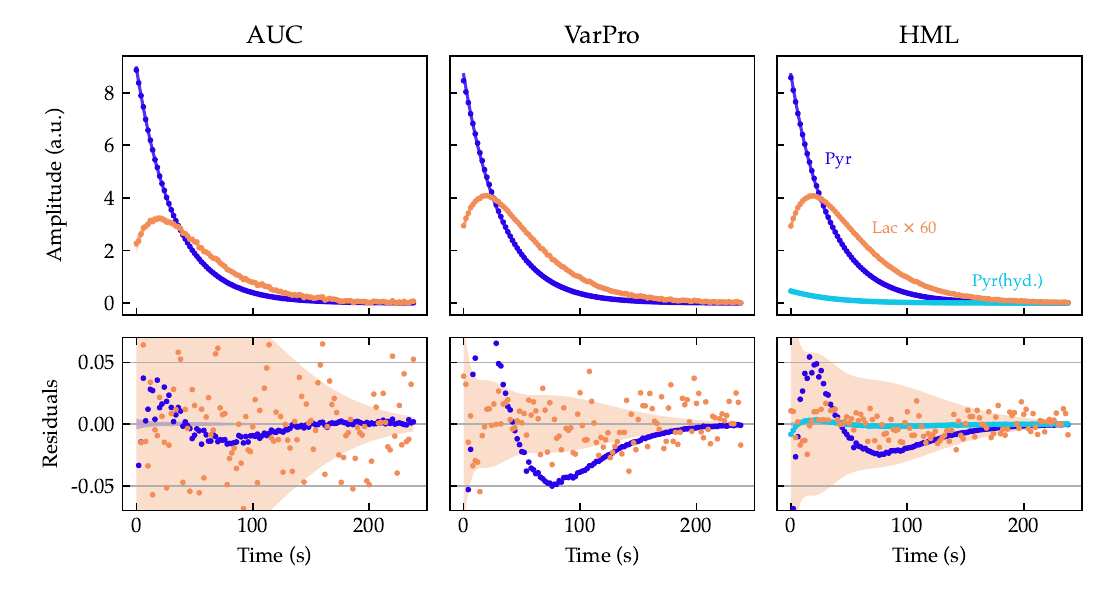}
	\caption{First experimental run: Signal curves (top plots) and residuals from the fit (bottom) for three different analysis approaches. Error bars and not drawn to declutter the image. The Lactate signal is also scale by a factor of \num{60} in the residual plots.}
	\label{fig:residuals_2}
\end{figure}

\begin{figure}
	\includegraphics[scale=1]{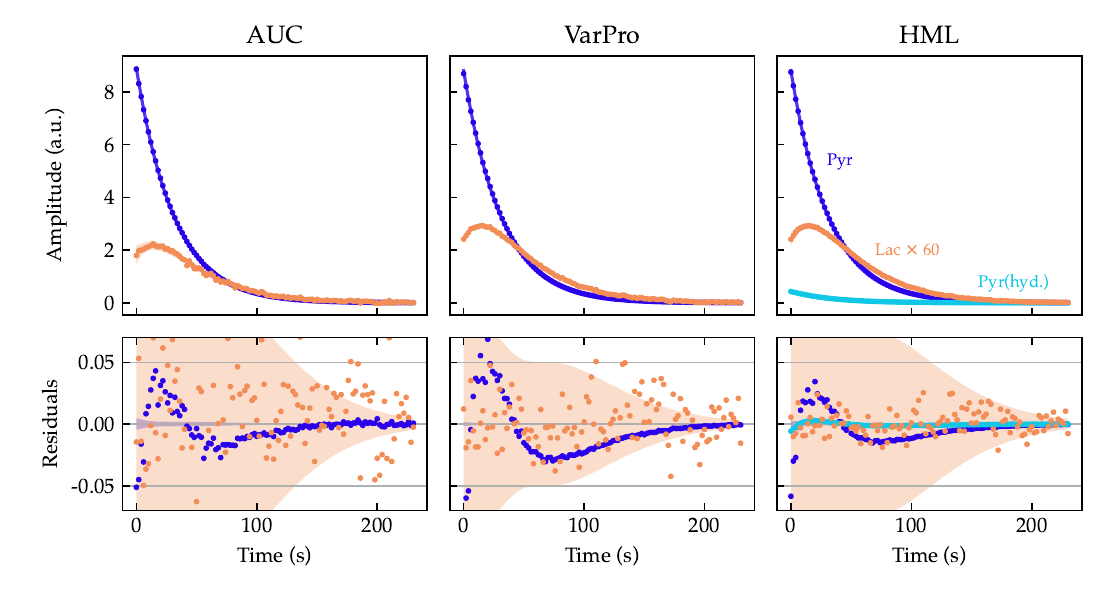}
	\caption{Second experimental run: Signal curves (top plots) and residuals from the fit (bottom) for three different analysis approaches. Error bars and not drawn to declutter the image. The Lactate signal is also scale by a factor of \num{60} in the residual plots.}
	\label{fig:residuals_3}
\end{figure}

\begin{figure}
	\includegraphics[scale=1]{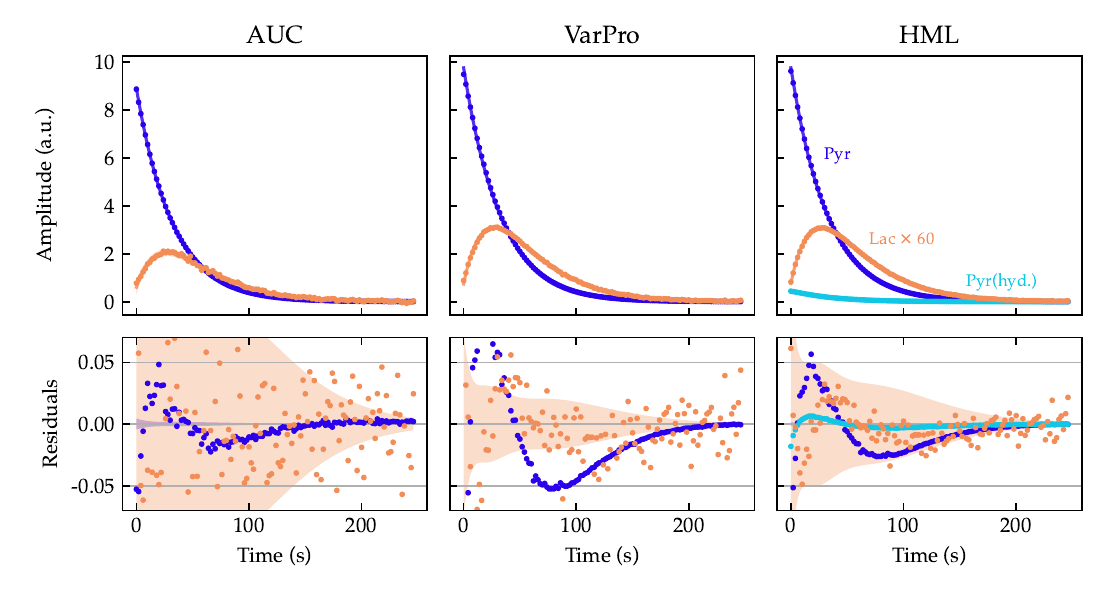}
	\caption{Third experimental run: Signal curves (top plots) and residuals from the fit (bottom) for three different analysis approaches. Error bars and not drawn to declutter the image. The Lactate signal is also scale by a factor of \num{60} in the residual plots.}
	\label{fig:residuals_4}
\end{figure}

The curve and residuals of the kinetic model fit using the data depicted in figs.~\ref{fig:data_correction_2}, \ref{fig:data_correction_3}, and \ref{fig:data_correction_4} are shown in figs.~\ref{fig:residuals_2}, \ref{fig:residuals_3}, and \ref{fig:residuals_4}, respectively.
Whereas for AUC and VarPro we do not necessarily have to also include the Pyruvate (hydrate) signal, the fit with the HML method requires inclusion of this resonance to provide an accurate description of the data.
All resonances show systematic deviation of the data from the used first-order kinetic model for the Pyruvate signal on the same order of magnitude for the three approaches and, thereby, validate each other.
The systematic deviation of the Lactate signal from the model, however, only really becomes apparent using the HML approach.

\section{2D FFT of the J-Coupling Spectroscopy using NV-Centers}
\begin{figure}
	\includegraphics[scale=1]{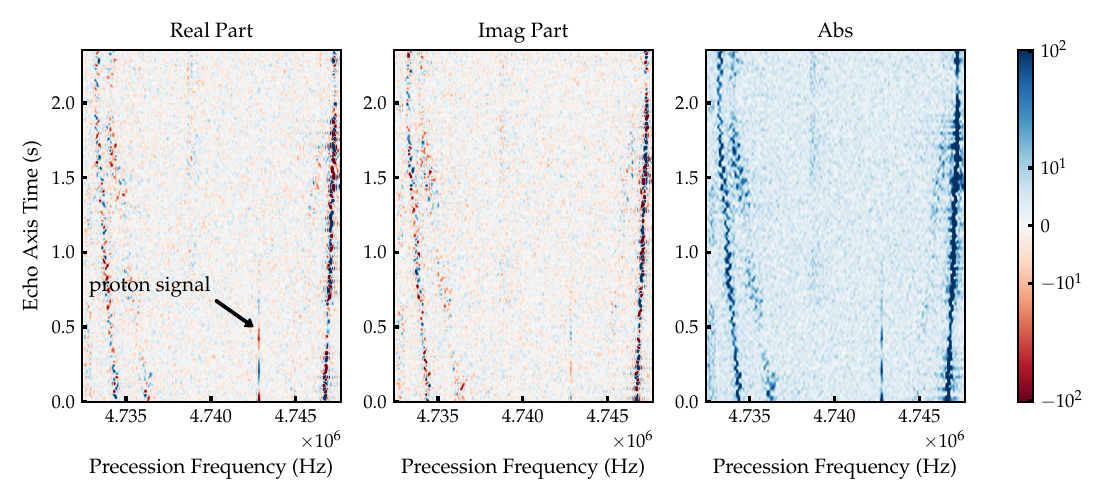}
	\caption{Signal of the J-coupling spectroscopy protocol on hyperpolarized Fumarate recorded with NV-centers over time. Each horizontal slab represents the FFT data.}
	\label{fig:fft_over_time}
\end{figure}

\begin{figure}
	\includegraphics[scale=1]{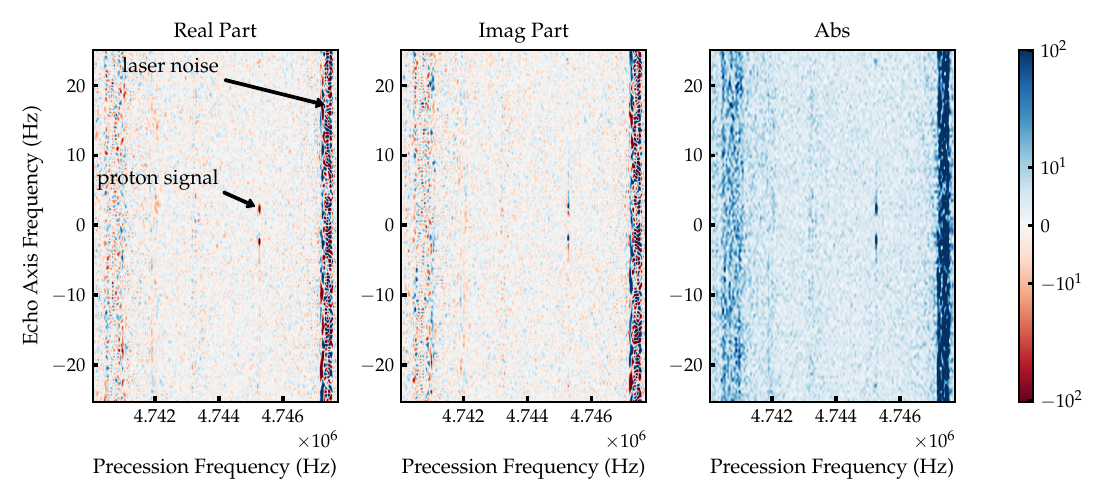}
	\caption{Signal of the J-coupling spectroscopy protocol on hyperpolarized Fumarate recorded with NV-centers after 2D FFT. The proton resonance is also indicated in fig.~4 of the main text.}
	\label{fig:2dfft}
\end{figure}

The J-coupling spectroscopy introduced in Section~3.2 of the main text records data of Larmor precession echoes with intermediate application of $\pi$-pulses on both spin species in \CFum{}.
The phase cycling applied in the protocol requires correction of the data by a complex phase factor before further processing.
When organized in a data matrix and upon application of FFT on the fast time scale of the Larmor precession makes the evolution due to the J-coupling visible in fig.~\ref{fig:fft_over_time} for the experimental run numbered "00083".
The dominant visible signal towards high frequencies is assumed to emerge from instability of the laser used for the read-out.
Upon application of the 2D-FFT as depicted in fig.~\ref{fig:2dfft} of the experimental run numbered "00083", one can also immediately extract the frequency in the J-evolution at around \qty{2.5}{\Hz}.

\end{document}